\documentclass[twocolumn]{aastex63}

\received{}
\revised{}
\accepted{}

\submitjournal{ApJ}

\shorttitle{Dust scale height}
\shortauthors{Doi and Kataoka}
\graphicspath{{./}{figures/}}
\begin{document}

\title{Estimate on dust scale height from \textcolor{black}{ALMA dust} continuum image of the HD 163296 protoplanetary disk}

\author[0000-0003-1958-6673]{Kiyoaki Doi}
\affiliation{Department of Astronomical Science, School of Physical Sciences, Graduate University for Advanced Studies (SOKENDAI), 2-21-1 Osawa, Mitaka, Tokyo 181-8588, Japan}
\affiliation{National Astronomical Observatory of Japan, 2-21-1 Osawa, Mitaka, Tokyo 181-8588, Japan}

\correspondingauthor{Kiyoaki Doi}
\email{doi.kiyoaki.astro@gmail.com}

\author[0000-0003-4562-4119]{Akimasa Kataoka}
\affiliation{National Astronomical Observatory of Japan, 2-21-1 Osawa, Mitaka, Tokyo 181-8588, Japan}

\nocollaboration{2}

\begin{abstract}
We aim at estimating the dust scale height of protoplanetary disks from millimeter continuum observations.
First, we present a general expression of intensity of a ring in a protoplanetary disk, and show that we can constrain the dust scale height by the azimuthal \textcolor{black}{intensity} variation.
Then, we apply the presented methodology to the two distinct rings at 68 au and at 100 au of the protoplanetary disk around HD 163296.
We constrain the dust scale height by comparing the DSHARP high-resolution millimeter dust continuum image with radiative transfer simulations using RADMC-3D.
We find that \textcolor{black}{$h_{\mathrm{d}}/h_{\mathrm{g}} > 0.84$} at the inner ring and \textcolor{black}{$h_{\mathrm{d}}/h_{\mathrm{g}} < 0.11$} at the outer ring with the \textcolor{black}{$3\sigma$} \textcolor{black}{uncertainties}, where $h_{\mathrm{d}}$ is the dust scale height and $h_{\mathrm{g}}$ is the gas scale height.
This indicates that the dust is flared at the inner ring and settled at the outer ring.
We further constrain the ratio of turbulence parameter $\alpha$ to gas-to-dust-coupling parameter $\mathrm{St}$ from the derived dust scale height; \textcolor{black}{$\alpha /  \mathrm{St} > 2.4$} at the inner ring, and \textcolor{black}{$\alpha /  \mathrm{St} < 1.1 \times 10^{-2}$} at the outer ring.
This result shows that the turbulence is stronger or the dust is smaller at the inner ring than at the outer ring.
\end{abstract}


\section{Introduction} \label{sec:intro}
Atacama Large Millimeter/submillimeter Array (ALMA) have changed the picture of protoplanetary disks, which are the fields of planet formation.
High spatial resolution observations with ALMA have revealed that many protoplanetary disks have structures in millimeter dust continuum such as spirals \citep{Perez2016}, crescents \citep{vanderMarel2013, Fukagawa2013, Casassus2015, Isella2018}, and rings \citep{alma2015, Fedele2017, Fedele2018, vanderPlas2017, Clarke2018, Dipierro2018, Andrews2018}.
The dust ring formation mechanism is still under discussion.
There are several possible scenarios for the dust ring formation;
dust accumulation by gas gaps induced by planets \citep[e.g.,][]{Lin1979, Goldreich1980, Zhu2012, Pinilla2012a, Pinilla2012b}, 
enhanced dust fragmentation by dust sintering near snowlines \citep{Okuzumi2016}, dust accumulation \textcolor{black}{at} the outer edge of the dead zone \citep{Gressel2015, Flock2015, Ueda2019}, and secular gravitational instability (SGI) \citep{Ward2000, Youdin2011, Michikoshi2012, Takahashi2014, Tominaga2019}.

The physical parameters of protoplanetary disks have been estimated by various methods.
The dust radius has been estimated by the spectral index of the dust emission \citep[e.g.,][]{Testi2014review}, and the gas surface density has been estimated by the line emission of gas \citep[e.g.,][]{Isella2016}.
The turbulence $\alpha$ is an important parameter: it controls the accretion of disks \textcolor{black}{\citep[e.g.,][]{Shakura1973, Lynden-Bell1974, Hartmann1998}, growth and fragmentation of the dust \citep[e.g.,][]{Blum_Muench1993_dust_collision_experiment, Blum2008ARAA, Brauer2008, Wada2013_dust_growth}, and mixing of the dust and the gas \citep[e.g.,][]{Dubrulle2005aa_dust_thickness, Youdin2007}.}
However, direct observation of the gas turbulence from line broadening has not been detected because of the low level of turbulence \citep[e.g.,][]{Dartois2003, Pietu2007AA_turbulence, Hughes2011_turbulence, Guilloteau2012turbulence, Flaherty2015, Flaherty2017, Teague2016, Teague2018} except for one tentative detection on DM Tau \citep{Flaherty2020}.

We focus on the dust scale height at the dust ring to constrain the physical state of protoplanetary disks.
\citet{Pinte2016} estimated the dust scale height of HL Tau \textcolor{black}{from the ring-gap contrast along the minor axis. }
The dust scale height \textcolor{black}{depends on} the ratio of dimensionless parameter of gas turbulence $\alpha$ \citep{Shakura1973} to dust-to-gas-coupling parameter called the Stokes number $ \mathrm{St} = \pi \rho_{\mathrm{mat}} a_{\mathrm{dust}} / 2 \Sigma_{\mathrm{gas}} $, where $\rho_{\mathrm{mat}}$ is dust material density, $a_{\mathrm{dust}}$ is dust radius, and $\Sigma_{\mathrm{gas}}$ is gas surface density \citep{Epstein1924, Adachi1976, Nakagawa1981, Nakagawa1986, Garaud2004, Dullemond2005, Fromang2006}.
Therefore, we can estimate the $\alpha/\mathrm{St}$ by estimating dust scale height.

In this paper, we estimate the dust scale height of HD 163296 from ALMA dust continuum observation.
The distance to this object is $101.5\ \mathrm{pc}$ \citep{Gaia2018}.
This object has two clear rings at 68 au and 100 au \citep{Isella2016, Andrews2018}.
There are indirect evidence of a planet at 260 au by kinematic signature \citep{Pinte2018} and at 78, 140, and 237 au by meridional flow \citep{Teague2019}.
\citet{Flaherty2015, Flaherty2017} put an upper limit of the gas turbulence that $\alpha \leq 3 \times 10^{-3}$.
\citet{Guidi2016} found a trend that the dust size decreases toward the outer region based on the spectral \textcolor{black}{index} at millimeter wavelength.
\citet{Dullemond2018} constrained $\alpha/\mathrm{St}$ based on the dust ring width and the upper limit of the gas ring width.
\citet{Rosotti2020} \textcolor{black}{expanded} \citet{Dullemond2018} and estimated $\alpha/\mathrm{St}$ by deviation of gas rotating velocity from the Keplerian rotation instead of the upper limit of the gas ring width.
\citet{Ohashi2019} estimated the dust scale height, the dust size, and the turbulence at the gaps by millimeter-wave polarization.
In this paper, we estimate the dust scale height, which becomes possible owing to the high spatial resolution by DSHARP (Disk Substructures at High Angular Resolution Project) campaign \citep{Andrews2018}.

The structure of this paper is as follows.
In section \ref{sec:analytical}, we consider a simple dust ring model to derive the general expression of the intensity as a function of the scale height.
We develop a method to constrain the dust scale height and consider the conditions under which the dust scale height can be determined from observations.
In section \ref{sec3}, we determine the dust scale height by applying the method to the image of HD 163296.
In section \ref{sec4}, we discuss physical quantities such as the turbulence and the dust size, and the ring formation mechanism based on the dust scale height.

\section{Analytical model} \label{sec:analytical}

In this paper, we aim at determining the dust scale height of dust rings in an inclined protoplanetary disk from \textcolor{black}{an} observed radio continuum image.
In this section, we assume a protoplanetary disk hosting an axisymmetric ring and analytically calculate the intensity of the disk.
We represent the dependence of the intensity on the dust scale height \textcolor{black}{if} the line of sight is inclined to the disk.
In section \ref{subsec2.1}, we describe the optical depth of the disk.
In section \ref{subsec2.2}, we calculate the intensity at the peak positions \textcolor{black}{along} the major and minor axes.
In section \ref{zyouken}, we discuss the conditions that allow us to determine the dust scale height.

\subsection{Optical depth of an inclined disk} \label{subsec2.1}
We consider the disk hosting an axisymmetric \textcolor{black}{dust} ring whose radial distribution of the dust surface density and the vertical distribution of the dust spatial density are described as Gaussian functions.
We take a cylindrical coordinate system as $r=0$ at the central star, $\phi=0$ along the major axis, and $z = 0$ at the midplane.
The surface density of the ring at the peak ($r=r_0$) is $\Sigma_0$, \textcolor{black}{the ring} width is $\sigma_r$, and \textcolor{black}{the scale} height is $\sigma_z$. 
The dust spatial density of this object is expressed as
\begin{equation}
\rho(r,z) = \frac{\Sigma_0}{\sqrt{2\pi} \sigma_{z}}\exp\left(-\frac{(r-r_{0})^{2}}{2\sigma_{r}^{2}} - \frac{z^{2}}{2 \sigma_{z}^{2}}\right).
\end{equation}

\begin{figure}[htbp]
    \begin{center}
      \includegraphics[width=8cm]{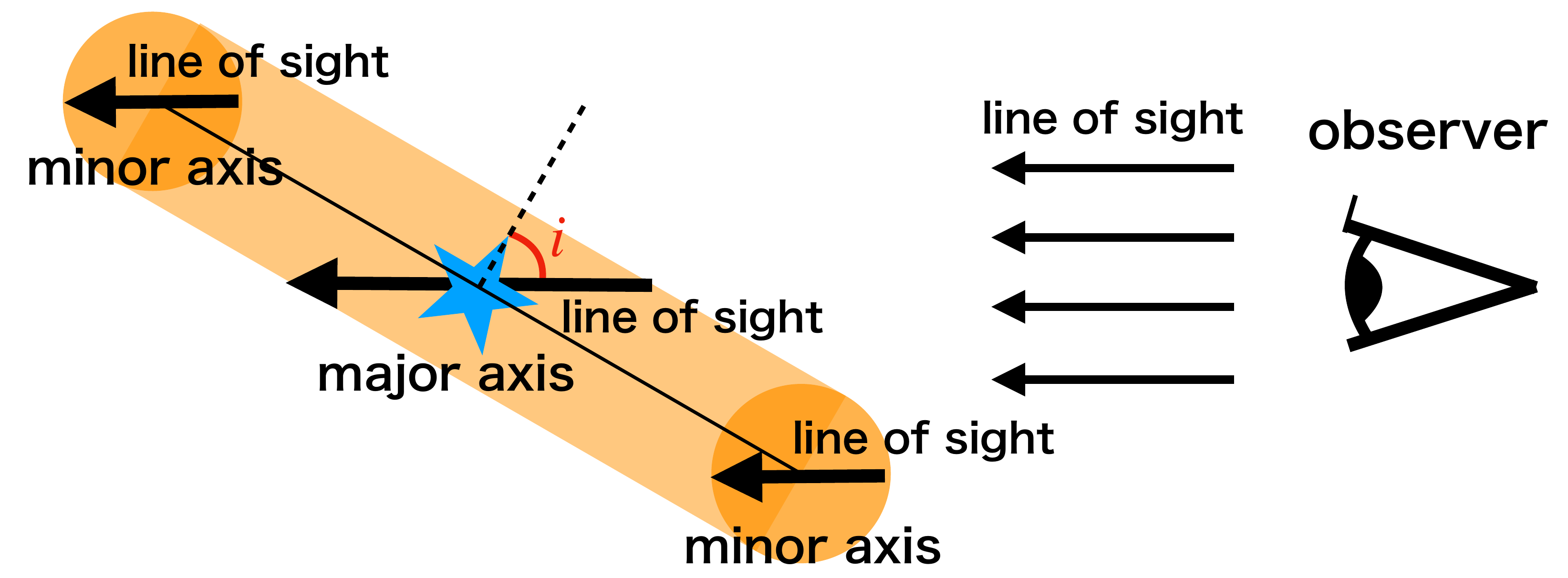}
      \caption{
This figure illustrates the principle that an optical depth is different on the major and minor axes if we observed it at an oblique angle.
}
    \label{danmen}
  \end{center}
\end{figure}

We assume the disk inclination angle to be $i$, and calculate the optical depth of this object along the line of sight. 
Figure \ref{danmen} illustrates the disk \textcolor{black}{geometry.}
In the calculation of the optical depth, we ignore the curvature of the ring. 
This approximation is appropriate \textcolor{black}{if} the ring radius is sufficiently larger than the ring width.
The optical depth at $r=r_{0}+\Delta r$ and $\phi$ is expressed as
\begin{eqnarray}
\tau(r_{0}+\Delta r,\phi) &=& \int_{-\infty}^{\infty} \kappa \rho(r_{0}+\Delta r + z \tan i \sin \phi,z)\frac{dz}{\cos{i}} \nonumber \\
&=& \frac{\kappa \Sigma_0}{\cos i} \frac{\sigma_r}{\sqrt{\sigma_r^2 + \sigma_z^2 \tan^2 i \sin^2 \phi}} \nonumber \\
&&\exp \left( -\frac{1}{2} \frac{\Delta r^2}{\sigma_r^2 + \sigma_z^2 \tan^2 i \sin^2 \phi}  \right) ,
\label{tau_ring}
\end{eqnarray}
where $\kappa$ is the dust opacity.

\textcolor{black}{The optical depth depends on the azimuthal angle, as shown in equation (\ref{tau_ring}). }
We use the word ``ridge'' to describe the peak position in the radial direction $(r=r_0)$ at each azimuthal angle.
In particular, the optical depth along the \textcolor{black}{ring's ridge} is the maximum on the major axis and \textcolor{black}{the} minimum on the minor axis.

\subsection{intensity on the major and minor axes} \label{subsec2.2}

\textcolor{black}{We calculate the azimuthal intensity variation of both optically thin and thick disks.}
Here, we ignore the effects of beam smearing in interferometric observations, which we discuss in Section \ref{zyouken}.

\textcolor{black}{
In the following discussion, we assume that the ring is isothermal in both vertical and radial directions.
In reality, there are temperature gradients in both the radial and directions.
We assume that the dust ring width is small, so the radial temperature variation can be ignored.
In the vertical direction, the temperature is higher in the upper layers that is irradiated by the central star \citep{Chiang1997, DAlessio1998, Dartois2003, Dalessio2005, Rosenfeld2013}.
For simplicity, we ignore the vertical temperature gradient.
We discuss the effect of the vertical temperature gradient in Appendix \ref{appendixA}.
}

For optically thick disks, the intensity can be approximated as
\begin{equation}
    I_{\nu}(r_0,\phi) = B_{\nu}(T) (1- \exp(-\tau)) \simeq B_{\nu}(T),
\end{equation}
where $T$ is the temperature of the \textcolor{black}{dust ring}, and $B_{\nu}(T)$ is the Planck function.
In other words, there is no variation of intensity if the ring is optically thick.

For optically thin disks, the intensity can be approximated as
\begin{equation}
    I_{\nu}(r_0,\phi) = B_{\nu}(T) (1- \exp(-\tau)) \simeq B_{\nu}(T) \tau(r_0,\phi).
\end{equation}
On the major axis($\phi = 0^\circ$), the intensity on the ridge is
\begin{equation}
    I_{\nu}(r_0,0^\circ) \simeq B_{\nu}(T) \frac{\kappa \Sigma_0}{\cos i},
    \label{major_axis_peak}
\end{equation}
and on the minor axis ($\phi = 90^\circ$), the intensity is
\begin{equation}
    I_{\nu}(r_0,90^\circ) \simeq B_{\nu}(T) \frac{\kappa \Sigma_0 \sigma_r}{\sqrt{\sigma_r^2 \cos^2i + \sigma_z^2 \sin^2i}}.
    \label{minor_axis_peak}
\end{equation}
The ratio of intensity on the ridge on the major axis to the minor axis is
\begin{equation}
    \frac{I_{\nu}(r_0,90^\circ)}{I_{\nu}(r_0,0^\circ)} = \frac{\sigma_r}{\sqrt{\sigma_r^2+\sigma_z^2 \tan^2{i}}}.
    \label{ratio_major_minor}
\end{equation}
Thus, there is a difference in intensity between the major and minor axes.

For the optically thin case, we further discuss \textcolor{black}{the difference of} the intensity between the major and minor axes.
As shown in equation (\ref{ratio_major_minor}), the projected dust scale \textcolor{black}{height, $\sigma_z \sin{i}$,} relative to the projected ring \textcolor{black}{width, $\sigma_r \cos{i}$,} is observed as the difference in the intensity between the major and minor axes.
If $\sigma_r \cos{i} \gg \sigma_{z} \sin{i}$, there is no difference in the intensity between the major and minor axes. 
On the other hand, \textcolor{black}{if} the projected ring \textcolor{black}{width, $\sigma_r \cos{i}$,} is not much wider than the projected dust scale \textcolor{black}{height, $\sigma_z \sin{i}$,} the difference in the intensity between the major and minor axes depends on the \textcolor{black}{dust} scale height.

\begin{figure*}[htbp]
    \begin{center}
      \includegraphics[width=18cm]{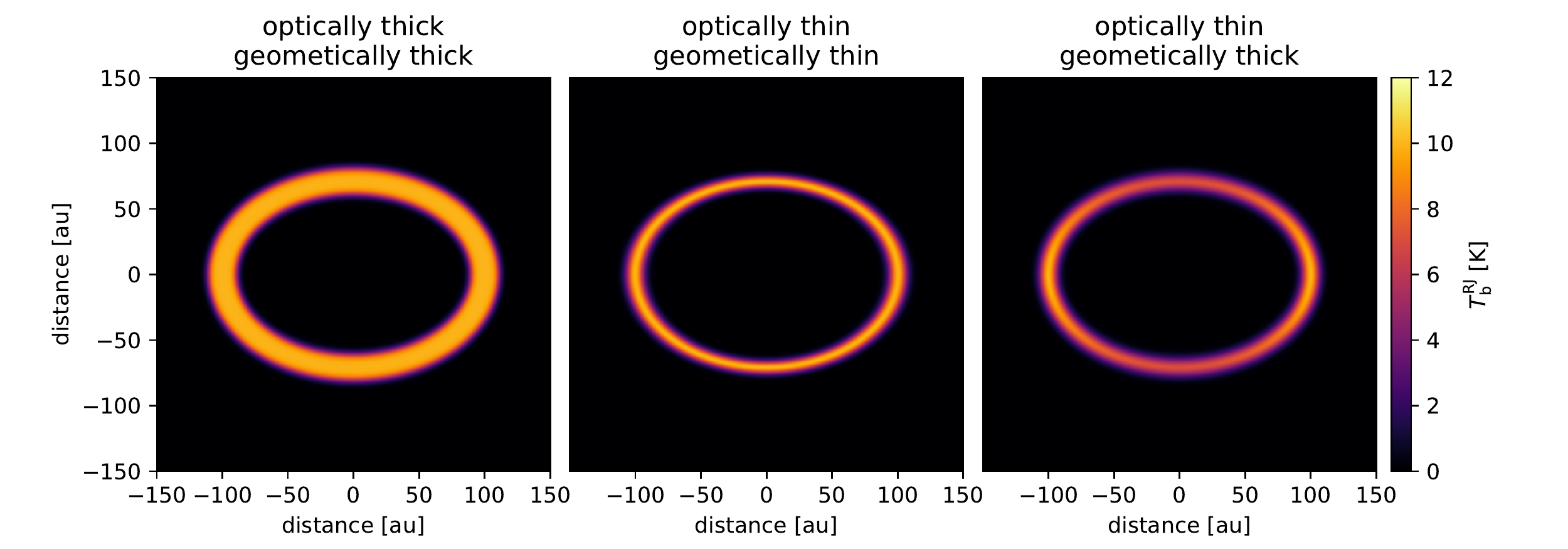}
      \caption{
Radiative transfer images of an optically thick and geometrically thick ring (left), an optically thin and geometrically thin ring (middle), and an optically thin and geometrically thick ring (right).
We assume that all of the rings are isothermal, and we set the temperature such that the brightness temperature at the peak position on the major axis is $10\ \mathrm{K}$. 
In the left figure, the optical depth at the \textcolor{black}{ring's} ridge is $\kappa \Sigma_0 = 5$, and the temperature is $15.0\ \mathrm{K}$. 
In the middle and right figures, the optical depth at the \textcolor{black}{ring's} ridge is $\kappa \Sigma_0 = 0.01$ and the temperature is $717.8 \mathrm{K}$.
The scale height of the middle figure is \textcolor{black}{5/16} au, and that of the left and right figures is \textcolor{black}{$5\ \mathrm{au}$}.}
      \label{model_image}
  \end{center}
\end{figure*}

\begin{figure}[htbp]
    \begin{center}
      \includegraphics[width=8cm]{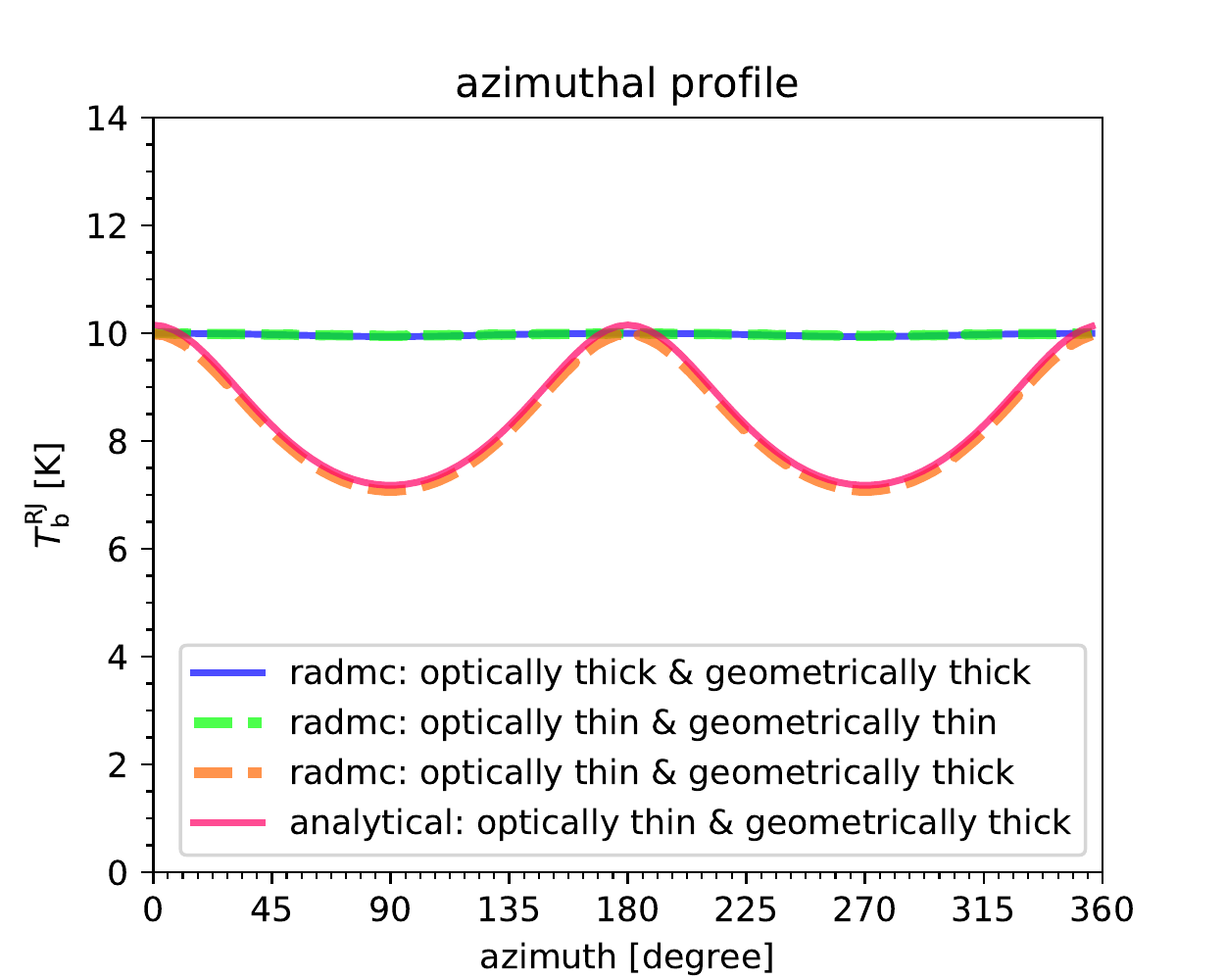}
      \caption{
Intensity along the ridges \textcolor{black}{in Figure \ref{model_image}}.
The blue, green, and orange lines represent the results of the simulation, and the pink line represents the analytical solution for the optically thin and geometrically thick case.}
    \label{model_ridge}
  \end{center}
\end{figure}

Figure \ref{model_image} shows images of radiative transfer simulations of protoplanetary disk models composed of a dust ring with different optical and geometric thicknesses.
We assume that all rings have the same width \textcolor{black}{($\sigma = 5\ \mathrm{au}$)} and distance from the center ($r_0 = 100\ \mathrm{au}$).
We assume that the rings are isothermal, and we set the temperature such that the brightness temperature at the peak position on the major axis is $10\ \mathrm{K}$. 
In the left figure, the optical depth \textcolor{black}{at the ring's ridge} is $\kappa \Sigma_0 = 5$, and the temperature is $15.0\ \mathrm{K}$. 
In the middle and right figures, the optical depth \textcolor{black}{at the ring's ridge} is $\kappa \Sigma_0 = 0.01$ and the temperature is $717.8 \mathrm{K}$.
The scale height \textcolor{black}{in} the middle figure is \textcolor{black}{5/16 au}, and that \textcolor{black}{in} the left and right figures is $5\ \mathrm{au}$.
Figure \ref{model_ridge} plots the intensity along the ridges of the simulated images shown in Figure \ref{model_image} and equation (\ref{eq:peak}) from the analytical calculations.
\textcolor{black}{We set the inclination to be $45^{\circ}$.}
We can see that for the optically thick (the left \textcolor{black}{panel}) or geometrically thin (the middle \textcolor{black}{panel}) cases, the intensity does not depend on the azimuthal direction, but for the optically thin and geometrically thick case (the right \textcolor{black}{panel}), the intensity on the minor axis is smaller than that on the major axis.
We confirm that the equation (\ref{eq:peak}) well describes the simulation results.
From the above, we conclude that the intensity along the ridges depends on the azimuthal angle if the ring is optically thin, the dust scale height is not too wide compared to the dust ring width, and the inclination is not too small.

\textcolor{black}{
Recent studies have pointed out that optically-thick emission can be fainter than the Planck function if the albedo is high \citep{Miyake1993, Liu2019scattering, Zhu2019scattering, Ueda2020, Lin2020scattering}. 
If the scattering works, one can imagine that the azimuthal variation may disappear due to the scattering-induced intensity reduction.
This does not change our conclusion if the ring is optically thin ($\tau_{\mathrm{abs}} \ll 1$) because the scattering effects are not significant. 
Moreover, this does not if the ring is extremely optically thick ($\tau_{\mathrm{abs}} \gg 1$) because we do not expect any azimuthal variation even if we do not consider the scattering effects. 
We should be careful if the ring is marginally optically thick ($\tau_{\mathrm{abs}} \sim 1$). 
In this case, if we do not consider the effects of scattering, we expect the azimuthal intensity variation to some extent. 
However, if scattering plays a role, while the intensity on both minor and major axes decreases, the intensity on the major axis decreases more than that on the minor axis. 
As a result, the azimuthal variation is suppressed compared with the non-scattering case, and it would be observed as if it is extremely optically thick. 
Therefore, the scattering effects may diminish the azimuthal variation of marginally optically thick rings. 
Note that the scattering may not significantly contribute to the emission in the case of HD 163296 because $\tau_{\mathrm{ext}} = \tau_{\mathrm{abs}} + \tau_{\mathrm{scat}} < 1$ from the observation of CO backside emission \citep{Isella2018}.
}
\subsection{Conditions for estimating dust scale heights from observations}\label{zyouken}

We discuss the conditions in which we can constrain the dust scale heights, including the effects of the beam smearing of the interferometric observations. 
We approximate the beam smearing as a Gaussian convolution of an ellipsoidal shape.
Based on the previous section, we only consider the optically thin case because we are not able to constrain the dust scale height \textcolor{black}{if} the dust ring is optically thick.
From equation (\ref{tau_ring}), the intensity at the location $(r+\Delta r, \phi)$ can be expressed as

\begin{eqnarray}
    I_{\nu}(r_0+\Delta r,\phi)
    &=& B_{\nu}(T)\frac{\kappa \Sigma_0}{\cos i} \frac{\sigma_r}{\sqrt{\sigma_r^2 + \sigma_z^2 \tan^2 i \sin^2 \phi}} \nonumber \\
    &&\exp \left( -\frac{1}{2} \frac{\Delta r^2}{\sigma_r^2 + \sigma_z^2 \tan^2 i \sin^2 \phi}  \right).
\end{eqnarray}

We convolve this equation with a Gaussian beam. 
We deproject the observed Gaussian beam for the inclination and denote the standard deviation of the deprojected beam along $\phi$ direction as $\sigma_{\mathrm{beam}} (\phi)$.
Here we ignore the curvature of the ring and convolve the radial profile and the beam smearing.
The convolved intensity is
\begin{eqnarray} \label{intensity_general}
    &&I_{\nu}(r_0+\Delta r,\phi) \nonumber
    \\
    &&= B_{\nu}(T)\frac{\kappa \Sigma_0}{\cos i} \frac{\sigma_r}{\sqrt{\sigma_r^2 + \sigma_z^2 \tan^2 i \sin^2 \phi + \sigma_{\mathrm{beam}}(\phi)^2}} \nonumber \\
    &&\exp \left( -\frac{1}{2} \frac{\Delta r^2}{\sigma_r^2 + \sigma_z^2 \tan^2 i \sin^2 \phi + \sigma_{\mathrm{beam}}(\phi)^2 }  \right) ,
\end{eqnarray}

Particularly, the intensities along the major axis and the minor axis are expressed as
\begin{eqnarray}
    I_{\nu}(r_0+\Delta r,0^{\circ})
    &=& B_{\nu}(T)\frac{\kappa \Sigma_0}{\cos i} \frac{\sigma_r}{\sqrt{\sigma_r^2 + \sigma_{\mathrm{beam}}(0^{\circ})^2}} \nonumber \\
    &&\exp \left( -\frac{1}{2} \frac{\Delta r^2}{\sigma_r^2 + \sigma_{\mathrm{beam}}(0^{\circ})^2}  \right),
    \label{eq:major axis}
\end{eqnarray}
\begin{eqnarray}
    I_{\nu}(r_0+\Delta r,90^{\circ}) && = B_{\nu}(T)\frac{\kappa \Sigma_0}{\cos i} \frac{\sigma_r}{\sqrt{\sigma_r^2 + \sigma_z^2 \tan^2 i + \sigma_{\mathrm{beam}}(90^{\circ})^2}} \nonumber \\
    && \exp \left( -\frac{1}{2} \frac{\Delta r^2}{\sigma_r^2 + \sigma_z^2 \tan^2 i + \sigma_{\mathrm{beam}}(90^{\circ})^2}  \right) .
    \label{eq:minor axis}
\end{eqnarray}
Along the \textcolor{black}{ring's} ridge, the intensity is expressed as
\begin{eqnarray}
    &&I_{\nu}(r_0,\phi) \nonumber \\
    &&= B_{\nu}(T)\frac{\kappa \Sigma_0}{\cos i} \frac{\sigma_r}{\sqrt{\sigma_r^2 + \sigma_z^2 \tan^2 i \sin^2 \phi + \sigma_{\mathrm{beam}}(\phi)^2}} .
    \label{eq:peak}
\end{eqnarray}

From equation (\ref{eq:major axis}), the intensity along the major axis is independent of the dust scale height. 
Therefore, we can construct a disk model that is independent of the dust scale height from the observed intensity along the major axis.
On the other hand, from equation (\ref{eq:minor axis}) and (\ref{eq:peak}), the intensity along the ridge except for the major axis or along the minor axis depends on the dust scale height of the ring. 
Therefore, we can constrain the dust scale height by the consistency of the observed intensity along the ridge or the minor axis.

We discuss the conditions for constraining the dust scale height from the observed images.
The dust scale height contributes to the intensity as the term $\sigma_z^2 \tan^2{i}$ in the denominator of equations (\ref{eq:minor axis}) and (\ref{eq:peak}).
If $\sigma_z^2 \tan^2{i}$ is not too small compared to the other terms ($\sigma^2_r$, $\sigma^2_{\mathrm{beam}}$), the effect of the dust scale height appears in the observed intensity.
Then, we can constrain the dust scale height from observations.
\textcolor{black}{We further discuss the effect of the beam smearing in Appendix \ref{appendix:beam}.}

To summarize the discussion above, we can constrain the dust scale height in the following steps.
First, we construct a disk surface density model from the observed intensity along the major axis.
Then, we search for the dust scale height that reproduces the observed intensity along the ridge or the minor axis.
We require the following four conditions to constrain dust scale heights:
1. dust ring is optically thin, 2. the dust ring width is not much wider than the dust scale height, 3. the inclination is not too small, and 4. the dust ring is spatially resolved.

\section{Application to HD 163296} \label{sec3}

We apply the method above to the protoplanetary disk around HD 163296 to constrain the dust scale height by comparing the observation and radiative transfer simulations.
In section \ref{subsec3.1}, we describe the observational data used in this work. 
In section \ref{subsec3.2}, we describe the disk model for the radiative transfer simulation. 
In section \ref{subsec3.3}, we compare the azimuthal variation of the intensity along the ridge as the first comparison method. 
In section \ref{subsec3.4}, we compare the radial \textcolor{black}{intensity} profile along the major and minor axes as the second comparison method.

\subsection{Observation} \label{subsec3.1}

In this study, we use the fits data of the image of the HD 163296 disk taken in the DSHARP campaign (2016.1.00484.L), which is one of the ALMA large programs \citep{Andrews2018}.
\textcolor{black}{This} image was taken with ALMA Band 6, which corresponds to the wavelengths of $\lambda = 1.25\ \mathrm{ mm}$. 
This image was made by combining the data of project 2013.1.00366.S13 \citep{Flaherty2015} and project 2013.1.00601.S4 \citep{Isella2016} to improve sensitivity and $uv$-coverage on shorter baselines. 
The details of this image are shown in \citet{Andrews2018} and \citet{Isella2018}. 
We assume that the distance to this object is $101.5\ \mathrm{pc}$ from Gaia DR2 parallax measurements \citep{Gaia2018}. 
We assume that the central star mass $M_{\mathrm{star}} = 2.02 M_{\mathrm{sun}}$ as a dynamical mass by line observation from \citet{Teague2019}.

\textcolor{black}{
This object satisfies the requirements for constraining the dust scale height described in section \ref{zyouken}.
The observation's beam size is $0\arcsec.020 \times 0\arcsec.016 $ with a standard deviation, which corresponds to $2.1 \times 1.6 $ au, and the beam position angle is $81.7^{\circ}$\citep{Isella2018}.
This source has narrow rings; the width is 4.0 au for the inner ring and 3.9 au for the outer ring with a standard deviation, which will be discussed in section \ref{subsec3.2}.
The disk inclination is $46.7^{\circ} \pm 0.1^{\circ}$, and the disk position angle is  $133.1^{\circ} \pm 0.1^{\circ}$ \citep{Isella2018}.
The gas scale heights of this source are 3.8 au in the inner ring and 6.7 au in the outer ring, using the temperature model discussed in section \ref{subsec3.2}.
As described above, this object is sufficiently inclined, and the ring width and beam are not too large compared to the gas scale height, so we think that we can constrain the dust scale height of this source.
}

\begin{figure}[htbp]
    \begin{center}
      \includegraphics[width=8cm]{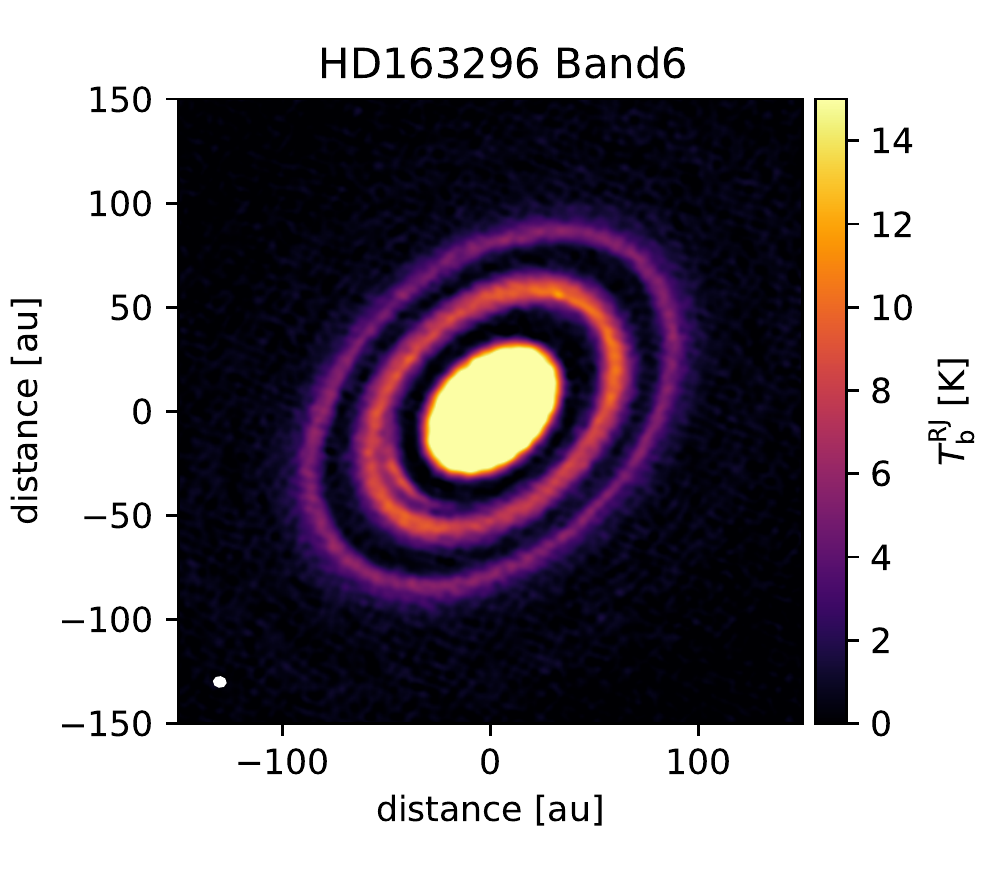}
      \caption{
Image of dust continuum of HD 163296 with ALMA Band 6 ($\lambda = 1.25\ \mathrm{mm}$). 
We use the brightness temperature, assuming Rayleigh-Jeans' law as a representation of intensity to preserve its linearity with the observed intensity.
}
    \label{observed_image}
  \end{center}
\end{figure}

\subsection{Setup of the simulation}  \label{subsec3.2}
\textcolor{black}{We} describe our simulation setup to compare with the observation.
We perform radiative transfer simulations with RADMC-3D \citep{Dullemond2012radmc}.
In our simulation, we use spherical coordinates $(r, \theta, \phi)$, and we assume axial symmetry in $\phi$ direction and plane symmetry with respect to the midplane.
The radial grid is linearly spaced between $r_{\mathrm{in}} = 10\mathrm{\ au}$ and $r_{\mathrm{out}} = 200\mathrm{\ au}$ with 512 grids.
The $\theta$ grid is linearly spaced with 512 grids in the range $0 \leq \theta \leq 0.2$, where $\theta = 0$ is the midplane, and $\theta = \pi/2$ is the north pole.

We adopt the following temperature profile with a smooth power-law distribution derived by \citet{Dullemond2020},
\begin{equation}
    T(r) = 18.7\ \mathrm{[K]} \left( \frac{r}{400\ \mathrm{[au]}} \right)^{-0.14}.
    \label{eq:temp}
\end{equation}
\citet{Dullemond2020} discussed the midplane temperature based on the CO line emission.
We assume that the disk is isothermal in the vertical direction.
In reality, the disk surface is hotter than the midplane, but gas vertical distribution is approximated well by hydrodynamic equilibrium at midplane temperature, so using the midplane temperature is a good approximation when we consider the gas vertical distribution.
\textcolor{black}{
We discuss the effect of the vertical temperature gradient on the intensity in Appendix \ref{appendixA}.
}

We use the dust opacity model of \citet{Birnstiel2018} that $\kappa_{\mathrm{abs}} = 0.484\ \mathrm{cm^2\ g^{-1}}$ at the wavelength of $\lambda = 1.25\ \mathrm{mm}$. 
Here, we consider only the absorption of the dust and do not include dust scattering \textcolor{black}{\citep[see][]{Miyake1993, Kataoka2014, Liu2019scattering, Zhu2019scattering, Ueda2020, Lin2020scattering}.}
We need dust surface density and dust opacity for the input parameters of RADMC-3D.
From the observed intensity and temperature distribution model of equation (\ref{eq:temp}), we can derive the optical depth but cannot solve the degeneracy of the dust surface density and dust opacity, so we need a dust opacity model to derive the dust surface density.
However, output images depend on the optical depth regardless of the combination of the dust opacity and surface density.
Therefore, our result does not depend on the dust opacity model.

We make a surface density profile based on the observational radial intensity profile along the major axis.
It is because the intensity along the major axis does not depend on the dust scale height, as shown in equation (\ref{eq:major axis}).
We use the radial intensity profile along the major axis in the northwest direction to avoid the additional crescent-shaped structure reported in the southeast direction \citep{Isella2018}.
We perform numerical simulations to search for parameters that reproduce the observational intensity along the major axis.
We assume that the dust surface density model is composed of an inner disk and two Gaussian rings as
\begin{eqnarray}
    \Sigma_{\mathrm{dust}}(r) &=& \Sigma_{\mathrm{disk}} \exp [ - ( r/r_{\mathrm{disk}} )^5 ] \nonumber \\
    &+& \Sigma_{\mathrm{ring1}} \exp [ - (r-r_{\mathrm{0,ring1}})^2/2 w_{\mathrm{ring1}}^2 ] \nonumber \\
    &+& \Sigma_{\mathrm{ring2}} \exp [ - (r-r_{\mathrm{0,ring2}})^2/2 w_{\mathrm{ring2}}^2 ].
\end{eqnarray}
We use the parameter that $\Sigma_{\mathrm{disk}}=20.66\ \mathrm{g\ cm^{-2}}$, $r_{\mathrm{disk}}=25.0\ \mathrm{au}$, $\Sigma_{\mathrm{ring1}}=1.26\ \mathrm{g\ cm^{-2}}$, $r_{\mathrm{0,ring1}}=67.9\ \mathrm{au}$, \textcolor{black}{$w_{\mathrm{ring1}}=4.0\ \mathrm{au}$, $\Sigma_{\mathrm{ring2}}=0.58\ \mathrm{g\ cm^{-2}}$, $r_{\mathrm{0,ring2}}=100.5\ \mathrm{au}$, and $w_{\mathrm{ring2}}=3.9\ \mathrm{au}$.}
In Section \ref{subsec3.4}, we discuss the consistency of \textcolor{black}{this model} and observation.
We summarize the parameters related to the rings, which we focus on in this study in Table \ref{tab:ringparameter}.
\textcolor{black}{
The absorption optical depth ($\tau_{\mathrm{abs}} = \kappa_{\mathrm{abs}} \Sigma_d / \cos{i}$) of this model at the rings is 0.89 for the inner ring and 0.41 for the outer ring.
\citet{Isella2018} estimated the extinction optical depth, $\tau_{\mathrm{ext}}$, of this source from the extinction of CO backside emission at the rings, and constrained that $\tau_{\mathrm{ext}} = 0.64\pm0.05$ for the inner ring and  $\tau_{\mathrm{ext}} = 0.74\pm0.05$ for the outer ring.
Since $\kappa_{\mathrm{ext}} = \kappa_{\mathrm{abs}} + \kappa_{\mathrm{scat}}$, $\tau_{\mathrm{ext}}$ must be larger than $\tau_{\mathrm{abs}}$.
However, $\tau_{\mathrm{abs}}$ in this study and constraint of $\tau_{\mathrm{ext}}$ in \citet{Isella2018} do not satisfy this relation for the inner ring.
A possible explanation for this inconsistency is that the dust surface density is overestimated because the temperature model used in this study underestimates the temperature, or $\tau_{\mathrm{ext}}$ in \citet{Isella2018} is underestimated due to beam smearing or contamination of front side emission.
}

\begin{deluxetable}{ccccc}
\tablecaption{Parameters of the dust rings of HD 163296\label{tab:ringparameter}}
\tablehead{
\colhead{}     & \colhead{$r_0$} & \colhead{$\Sigma_\mathrm{dust}$}  & \colhead{$w_{\mathrm{dust}}$} & \colhead{$h_{\mathrm{gas}}$}\\
\colhead{Ring} & \colhead{(au)}  & \colhead{($\mathrm{g\ cm^{-2}}$)} & \colhead{(au)} & \colhead{(au)} \\
\colhead{(1)} & \colhead{(2)} & \colhead{(3)} & \colhead{(4)} & \colhead{(5)} 
}
\startdata
inner ring & 67.9  & 1.26 & \textcolor{black}{4.0} & \textcolor{black}{3.8} \\
outer ring & 100.5 & 0.58 & 3.9 & 6.7 
\enddata
\tablecomments{(2) Distance to the central star. (3) The dust surface density at the peak of the rings. (4) The width of the dust rings. (5) Gas scale height based on midplane temperature and central star mass.}
\end{deluxetable}

Now we are ready to determine the dust scale height.
To estimate the scale height that reproduces the observation, we change the dust settling parameter $f_{\mathrm{set}}$. 
We define the dust settling parameter such that $f_{\mathrm{set}} = h_{\mathrm{gas}}/h_{\mathrm{dust}}$, where $h_{\mathrm{dust}}$ is dust scale height and $h_{\mathrm{gas}}$ is gas scale height.
The gas pressure scale height $h_{\mathrm{gas}} = c_s/\Omega_{\mathrm{kep}}$ from the hydrostatic equilibrium where $c_s$ is sound speed and $\Omega_{\mathrm{kep}}$ is the Keplerian orbital velocity. 
\textcolor{black}{We} run multiple simulations in the range of $1/2 \leq f_{\mathrm{set}} \leq 16$ with every $2^{1/16}$ steps, which corresponds to 81 models.
\textcolor{black}{
We convolve the simulated images with the beam and compare them with the observed image.
}

\subsection{Comparison of the intensity: The azimuthal variation along the ridge}  \label{subsec3.3}

\begin{figure*}[thbp]
    \begin{center}
      \includegraphics[width=18cm]{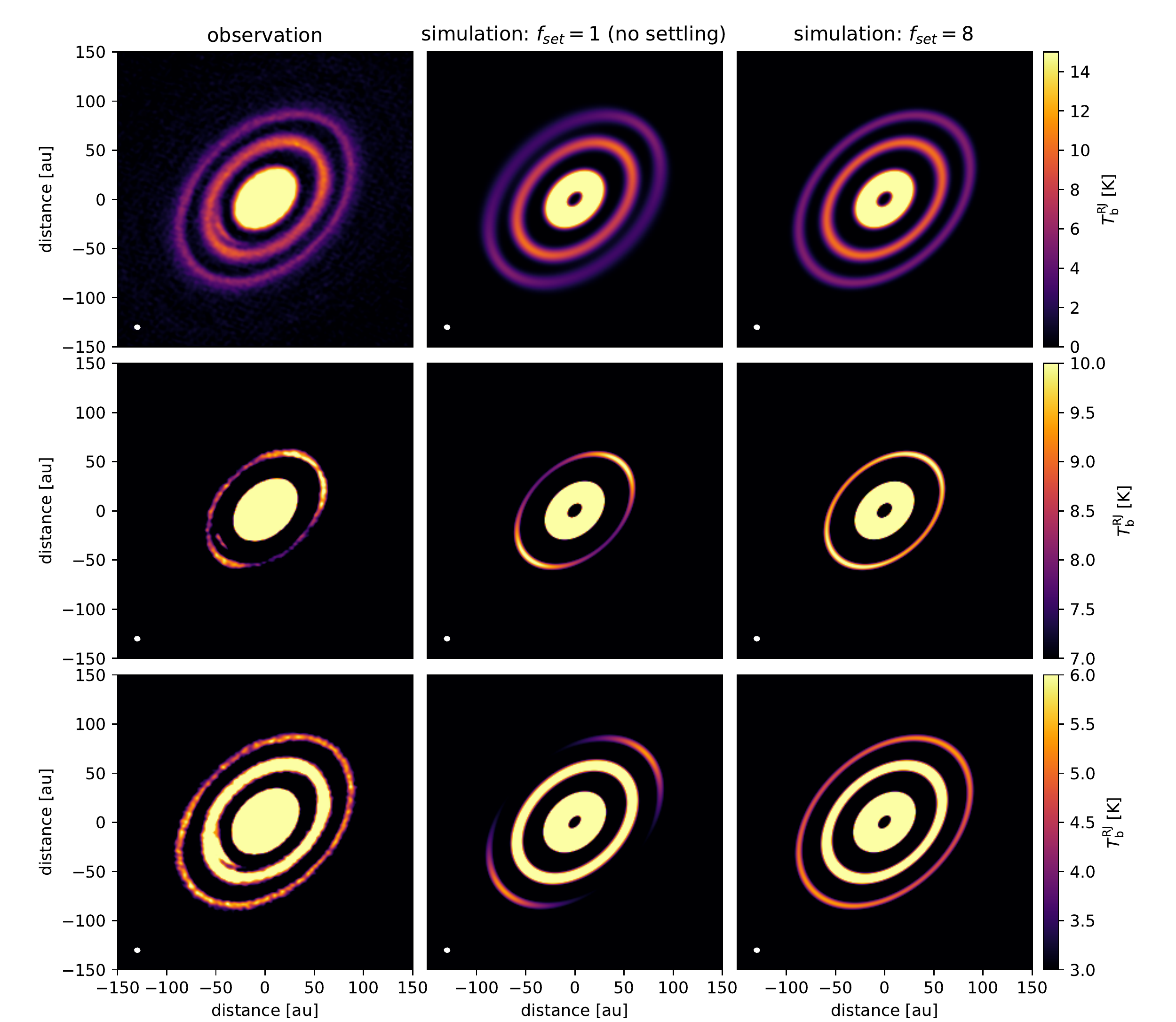}
      \caption{
Comparison of images of HD 163296. 
The three images in the left column represent observation, and the three images in the middle and right columns represent simulations with $f_{\mathrm{set}}=1$ (no settling) and $f_{\mathrm{set}}=8$, respectively.
Each line shows the images with different color ranges.
The middle line shows the \textcolor{black}{images that emphasize} the azimuthal \textcolor{black}{intensity} variation of the inner ring, and the bottom line shows that of the outer ring.
} 
      \label{manyfigure}
  \end{center}
\end{figure*}

The analytical expression of the azimuthal variation described as equation (\ref{eq:peak}) shows that the intensity along the ridge depends on the dust scale height.
We constrain the dust scale height by comparing the azimuthal \textcolor{black}{intensity} variation along the ridge of observation \textcolor{black}{with that of the} simulations.

\begin{figure}[htbp]
    \begin{center}
      \includegraphics[width=8.0cm]{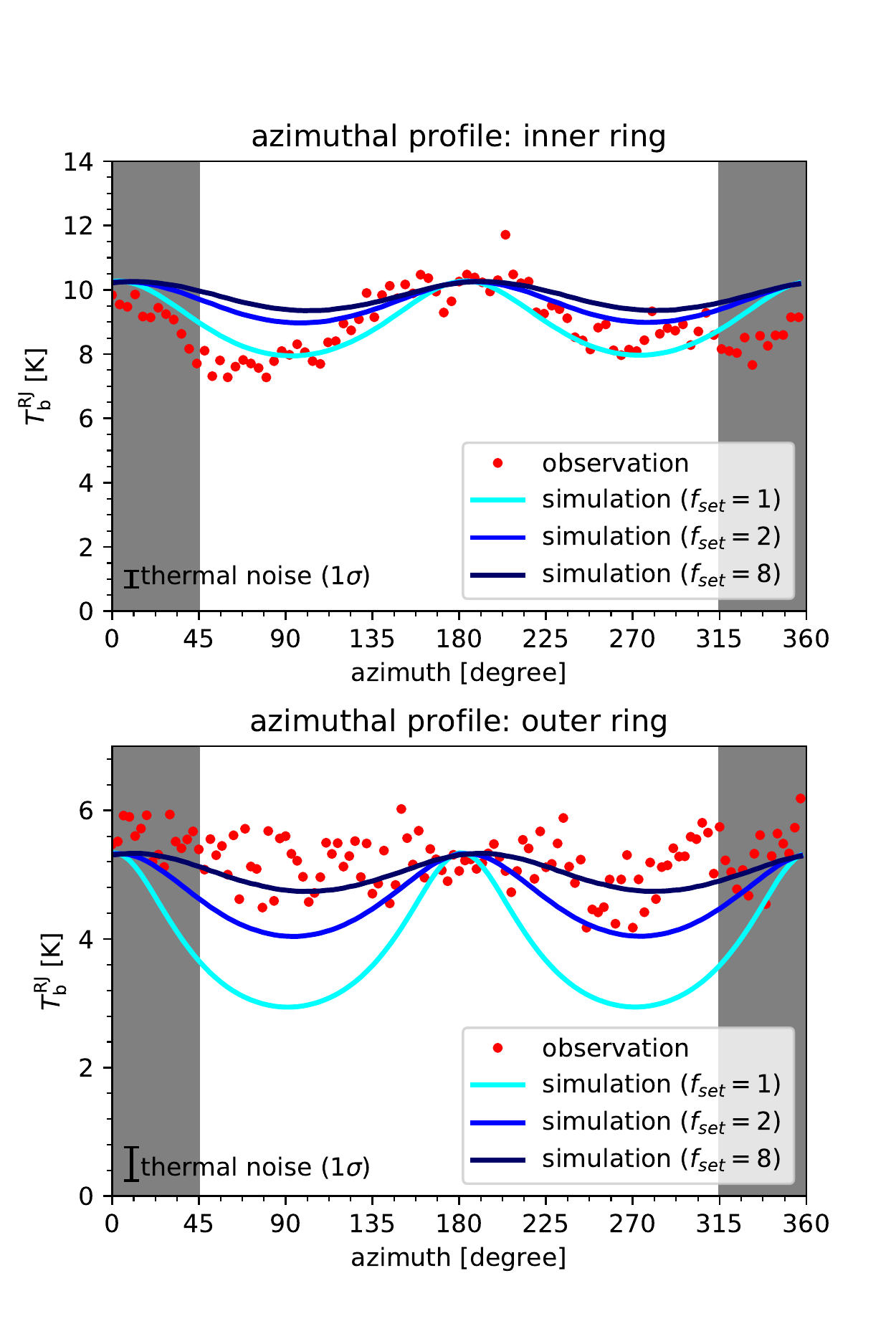}
      \caption{
Azimuthal \textcolor{black}{intensity} variation along the ridge.
The red dots mean \textcolor{black}{the} observation, and the solid lines mean \textcolor{black}{the} simulations. 
The black, blue, and cyan line mean $f_{\mathrm{set}}=1$ (no settling), ${2}$ and ${8}$, respectively. 
The shaded regions represent the area which have non-axisymmetric features like a crescent.
We plot the data at every FWHM of the beam \citep{Huang2018}.
}
      \label{degree-BT}
  \end{center}
\end{figure}

To demonstrate the azimuthal \textcolor{black}{intensity} variation, in Figure \ref{manyfigure}, we show images of the observation and simulations in the case of $f_{\mathrm{set}}= 1$ (no settling) and 8 (settled).
We use a different color range for each row to emphasize the azimuthal variations of the inner and outer rings.
Figure \ref{degree-BT} shows the azimuthal variation along the ridges of observation and simulations in the case of $f_{\mathrm{set}}= 1, 2,\ \mathrm{and} \ 8$.
As shown in the simulation images of $f_{\mathrm{set}} = 1$, where the dust is flared, the intensity along the ridge is larger on the major axis ($\phi = 0^\circ, 180^\circ$) than on the minor axis ($\phi = 90^\circ, 270^\circ$).
In contrast, as shown in  the simulation images of $f_{\mathrm{set}} = 8$, where the dust is settled, the intensity along the ridge does not depend on the azimuthal angle.
The observational images show that the intensity along the ridge of the inner ring is lower on the minor axis ($\phi = 90^\circ, 270^\circ$) than on the major axis ($\phi = 0^\circ, 180^\circ$).
In contrast, the intensity along the ridge of the outer ring is uniform for the azimuthal direction.
It infers that the dust scale height is large for the inner ring and small for the outer ring.

\begin{figure}[htbp]
    \begin{center}
      \includegraphics[width=7.0cm]{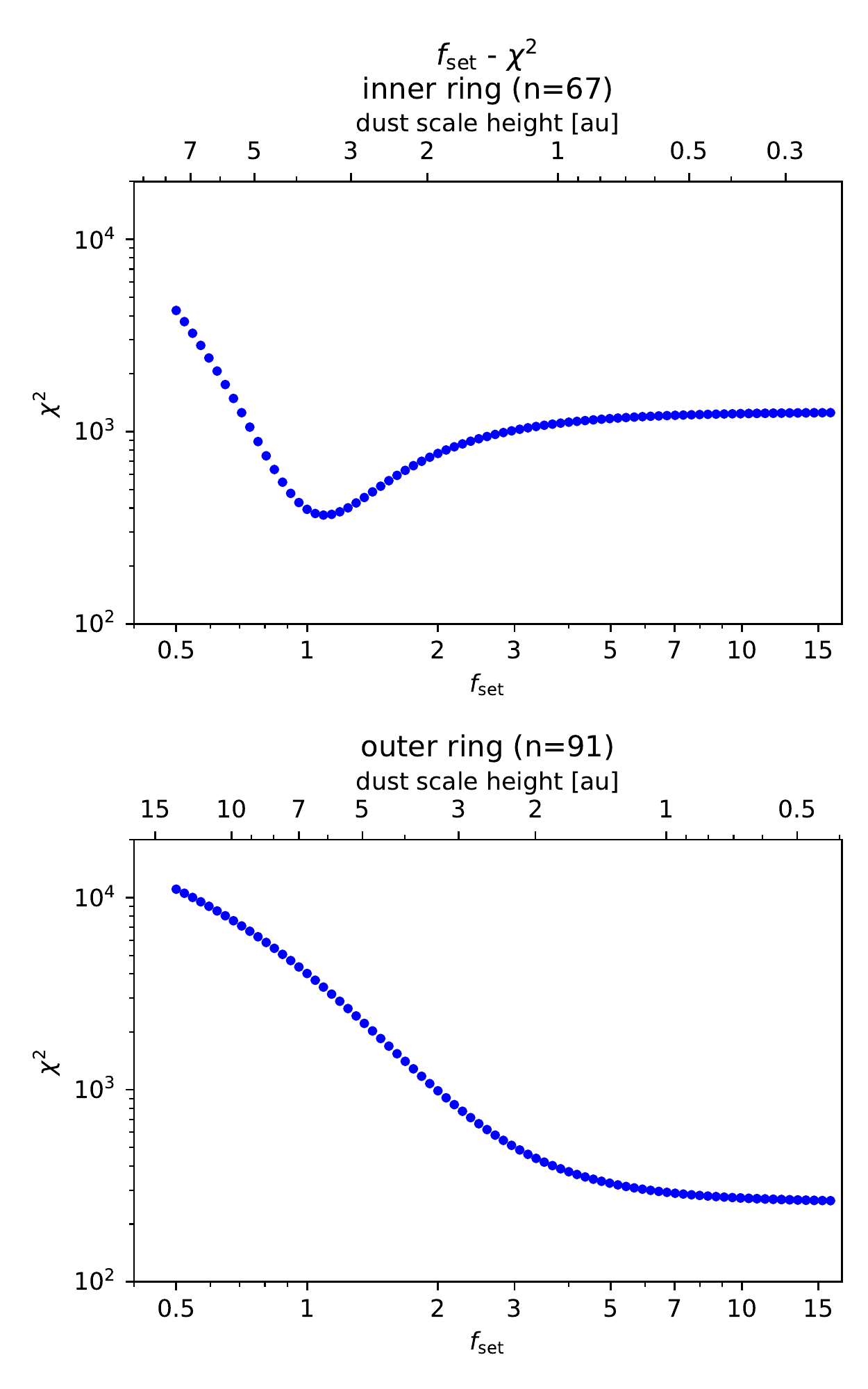}
      \caption{
\textcolor{black}{$\chi^2$} of the intensity between \textcolor{black}{the} observation and simulations along the ridge for each $f_{\mathrm{set}}$.
We use data at every FWHM.
We exclude data for the \textcolor{black}{region} which has a crescent structure (shaded area in Figure \ref{degree-BT}).
}
      \label{chi2}
  \end{center}
\end{figure}

We constrain $f_{\mathrm{set}}$ using \textcolor{black}{$\chi^2$}.
We run simulations varying $f_\mathrm{set}$ from 1/2 to 16 by a factor of $2^{1/16}$, and calculate \textcolor{black}{$\chi^2$}.
To estimate the parameter $f_{\mathrm{set}}$, we calculate \textcolor{black}{$\chi^2$} between the observation and the simulations along the ridge.
To exclude the effect of the crescent-shaped substructure in the southeast direction, we excluded the region between $-45^{\circ}$ and $45^{\circ}$ in our calculations of the \textcolor{black}{$\chi^2$}.
Figure \ref{chi2} shows the \textcolor{black}{$\chi^2$} at each $f_{\mathrm{set}}$.
The best-fit parameters are $f_{\mathrm{set}} = 1.1$ for the inner ring and $f_{\mathrm{set}} = 16$ for the outer ring.
\textcolor{black}{We calculate the range where $\chi^2 < \chi^2_{\mathrm{min}} + 3^2$,} and we obtain \textcolor{black}{$f_{\mathrm{set}} =1.1 ^{+0.1}_{-0.1}$ for the inner ring and  $f_{\mathrm{set}} > 9.5$} for the outer ring with $3\sigma$ uncertainty.
\textcolor{black}{It means that} the inner ring is more flared than the outer \textcolor{black}{ring.}
Figure \ref{chi2} also shows that the sensitivity to estimate $ f_{\mathrm{set}}$ is high in the region where $ f_{\mathrm{set}}$ is small, while the sensitivity is low in the region where $ f_{\mathrm{set}}$ is \textcolor{black}{large.}
\textcolor{black}{Note that $f_\mathrm{set}$ cannot be smaller than unity because the dust scale height cannot be larger than the gas scale height, while our results satisfies this condition.}


\subsection{Consistency with the radial profiles}   \label{subsec3.4}

We discuss whether the derived dust scale height from the azimuthal intensity variation is consistent with the observed radial intensity profile along the minor axis.
From equations (\ref{eq:major axis}) and (\ref{eq:minor axis}), the intensity along the major axis does not depend on the dust scale height, while that along the minor axis depends on the \textcolor{black}{dust} scale height. 
Along the minor axis, the higher the dust scale height is, the wider the observed dust ring width is, and the lower the observed intensity at the peak is.
We plot the intensity of both simulations and the observation along the major and minor axes to check the consistency of $f_{\mathrm{set}}$, which we estimate in section \ref{subsec3.3}.

\begin{figure}[htbp]
    \begin{center}
      \includegraphics[width=8cm]{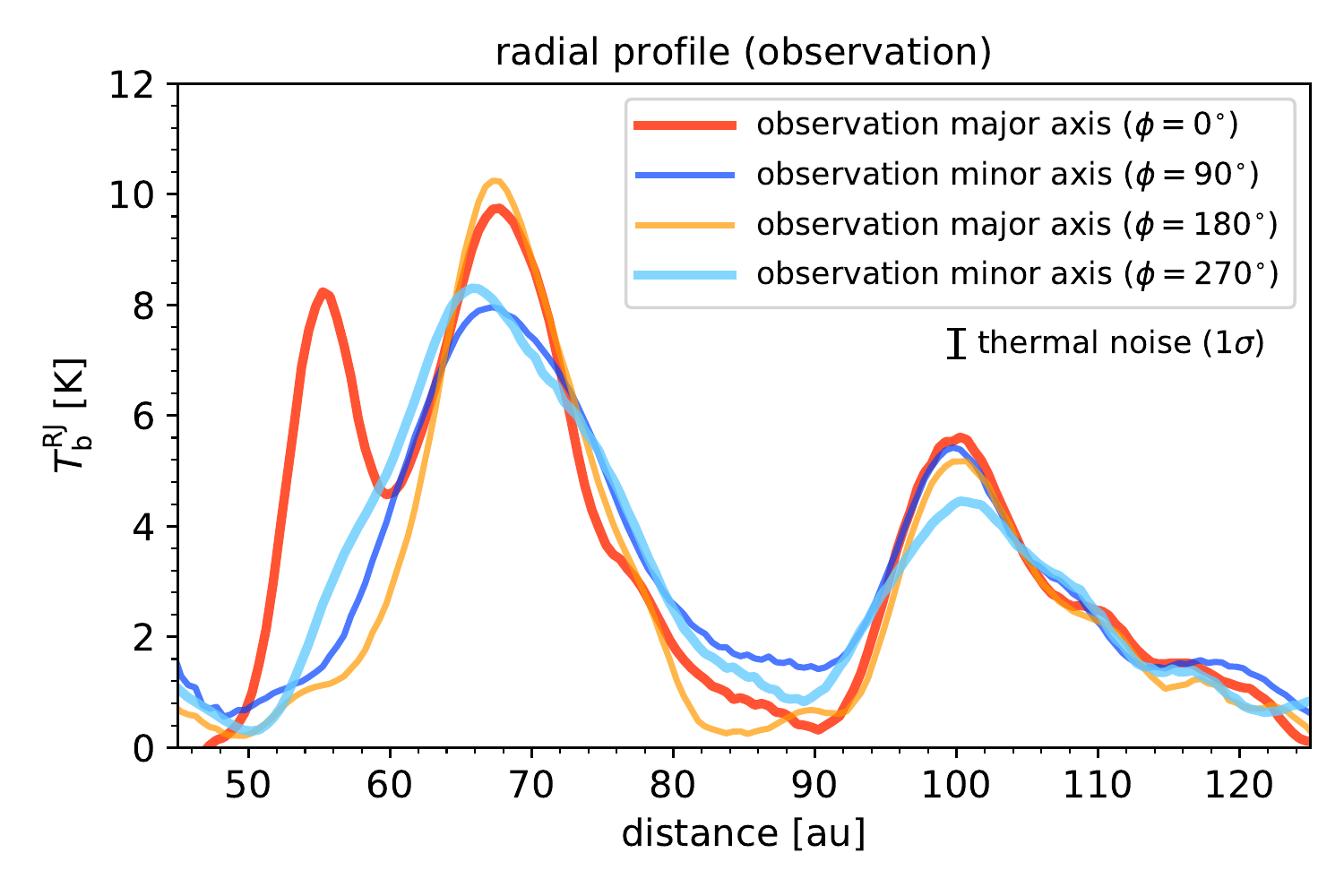}
      \caption{
The radial \textcolor{black}{intensity} profiles of \textcolor{black}{the} observation, which are averaged over 6 degrees width from the major and minor axes.
The blue and purple lines mean intensity along the major axis, and the orange and red ones mean that along the minor axis. 
}
      \label{observation_radial}
  \end{center}
\end{figure}

Figure \ref{observation_radial} shows the radial \textcolor{black}{intensity} profiles of the observation along the major and minor axes.
Around the inner ring at 68 au, the intensity along the major axis is larger than that along the minor axis.
On the other hand, around the outer ring at 100 au, the intensity along the major and minor axes is almost the same.

\begin{figure}[htbp]
    \begin{center}
      \includegraphics[width=8cm]{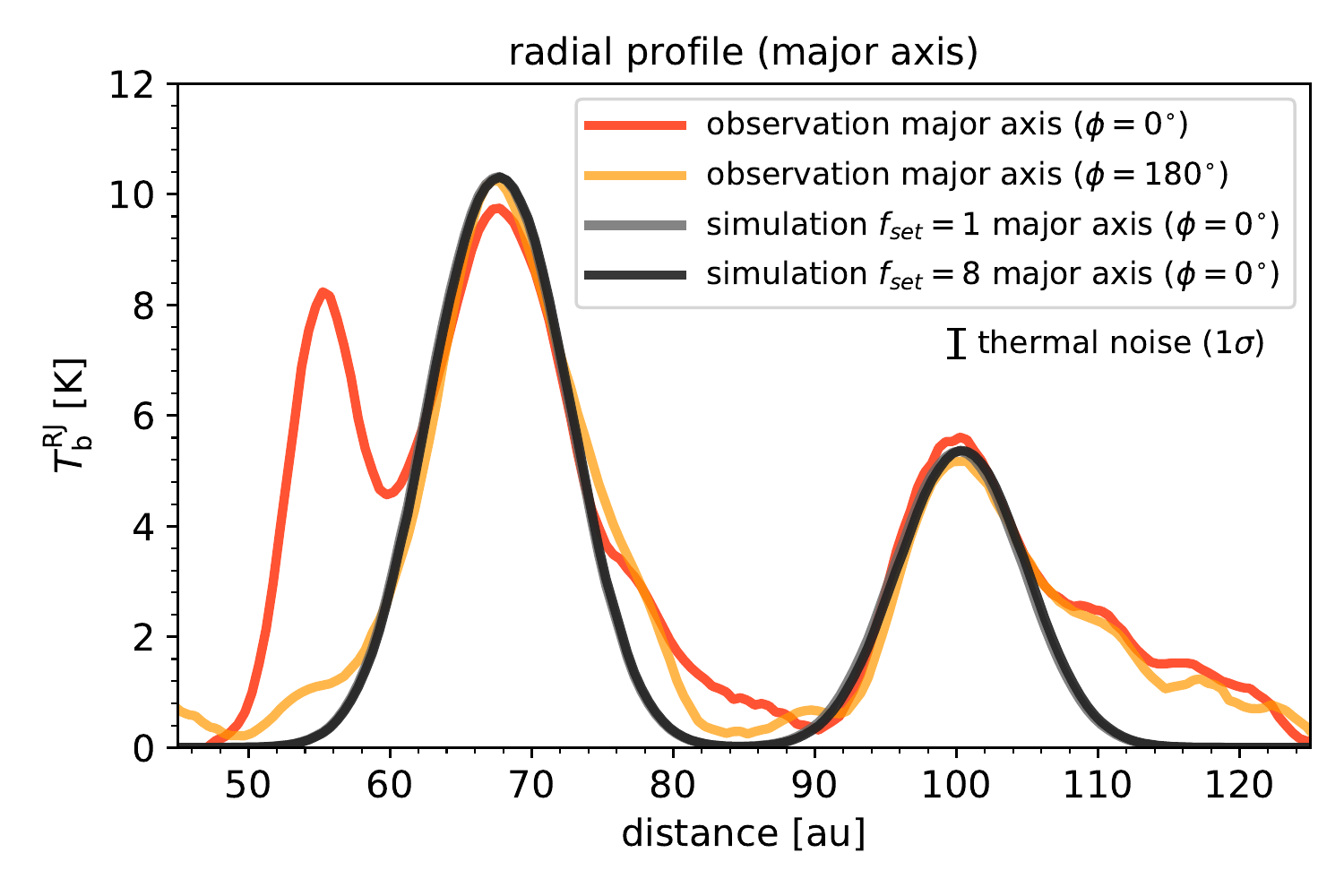}
      \includegraphics[width=8cm]{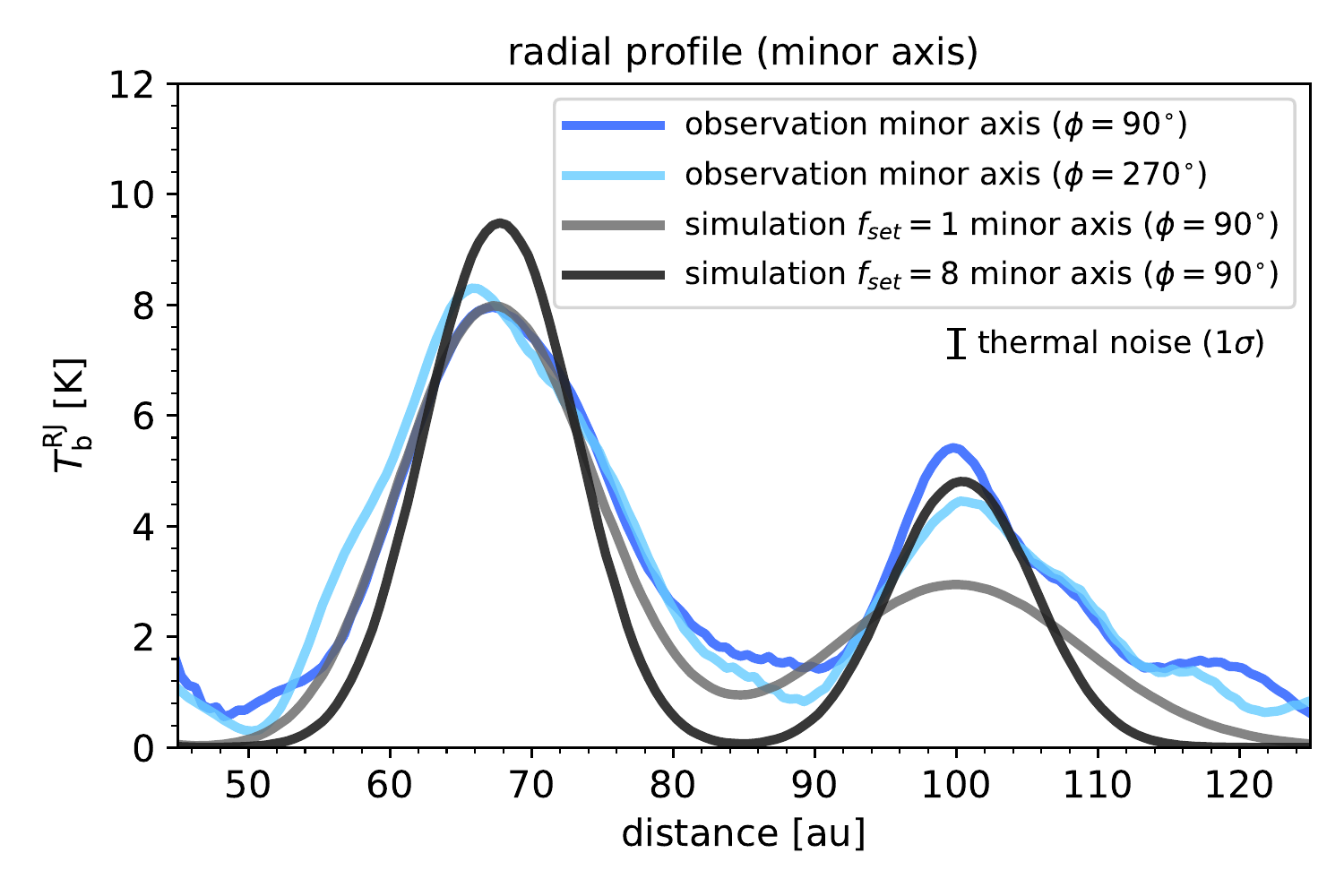}
      \caption{
The radial \textcolor{black}{intensity} profiles along the major axis (top) and minor axis (bottom). 
The black and gray lines are the intensity of \textcolor{black}{the} simulations, and the other lines are the intensity
of observation. 
The intensity \textcolor{black}{along} the major axis does not depend on the dust scale height, so the two simulation lines overlap.
}
    \label{radial2fig}
  \end{center}
\end{figure}

Figure \ref{radial2fig} shows the radial intensity profiles along the major (top) and minor (bottom) axes of the observation and simulations in case of $f_{\mathrm{set}}= 1\ \mathrm{and}\ 8$.
We can see that the simulations reproduces the observation well \textcolor{black}{along} the major axis \textcolor{black}{from the upper panel of Figure \ref{radial2fig}.}
\textcolor{black}{The bottom panel of} Figure \ref{radial2fig} shows that the radial profile along the minor axis around the inner ring is well reproduced by $f_{\mathrm{set }}=1$, while that around the outer ring is well reproduced by $f_{\mathrm{set}}=8$.
This result is consistent with \textcolor{black}{the result in section \ref{subsec3.3}} that the inner ring is flared, and the outer ring is settled.

\begin{figure}[htbp]
    \begin{center}
      \includegraphics[width=8cm]{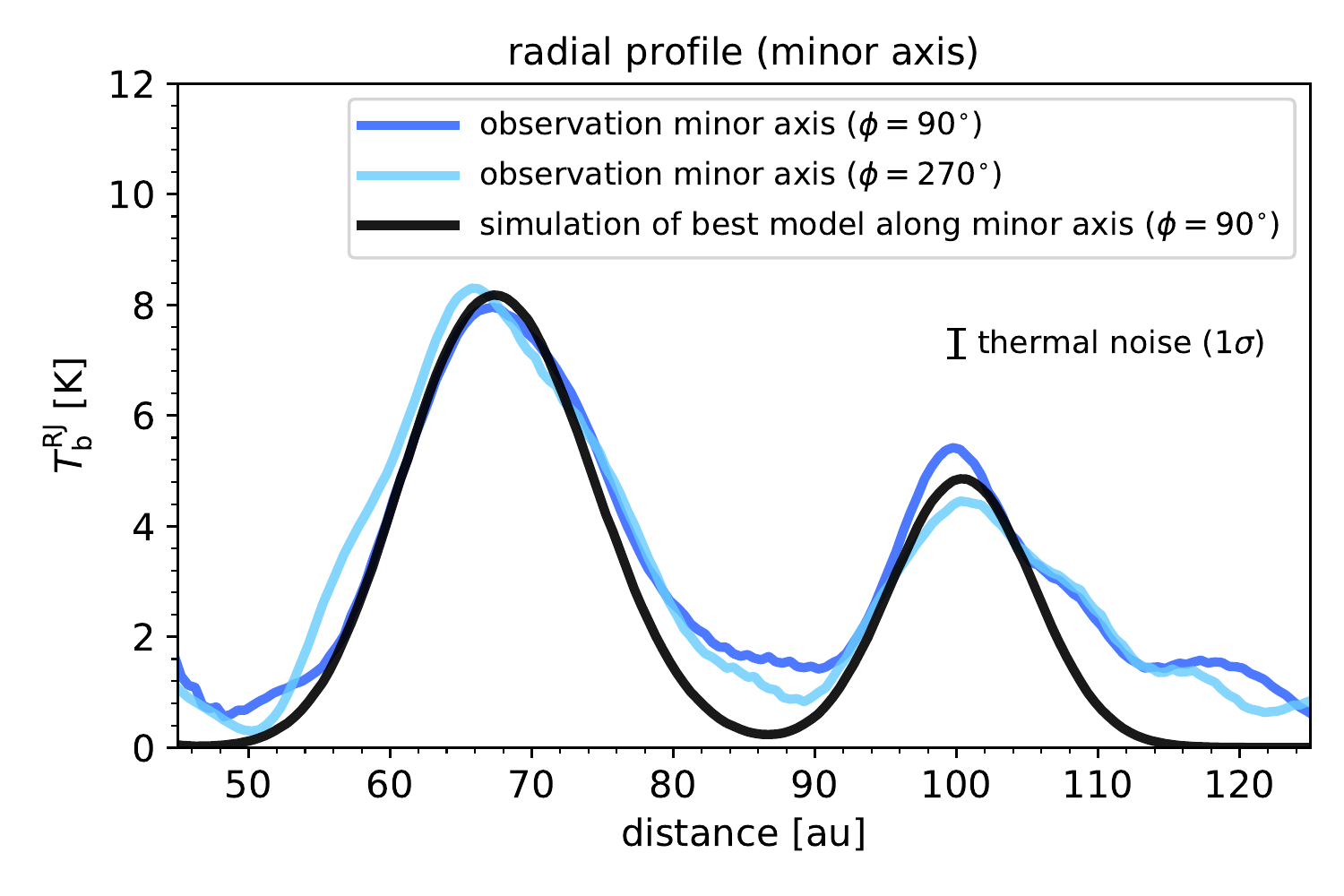}
      \caption{
The radial \textcolor{black}{intensity} profiles along the minor axis. 
The black line represents the intensity of the best model ($f_{\mathrm{set}}=1.1$ for the inner ring and $f_{\mathrm{set}} = 16$ for the outer ring), and the blue and cyan lines are that of \textcolor{black}{the} observation.
}
      \label{minor_axis_radial_best}
  \end{center}
\end{figure}

Figure \ref{minor_axis_radial_best} \textcolor{black}{shows} the radial \textcolor{black}{intensity} profile of the best model along the minor axis.
Based on the result \textcolor{black}{in section} \ref{subsec3.3}, we set $f_{\mathrm{set}}=1.1$ for the inner ring and $f_{\mathrm{set}}=16$ for the outer ring.
This result reproduces the onservation well for the entire region.

\section{discussion} \label{sec4}

\subsection{The dust scale height for the entire region}
\begin{figure}[htbp]
    \begin{center}
      \includegraphics[width=9cm]{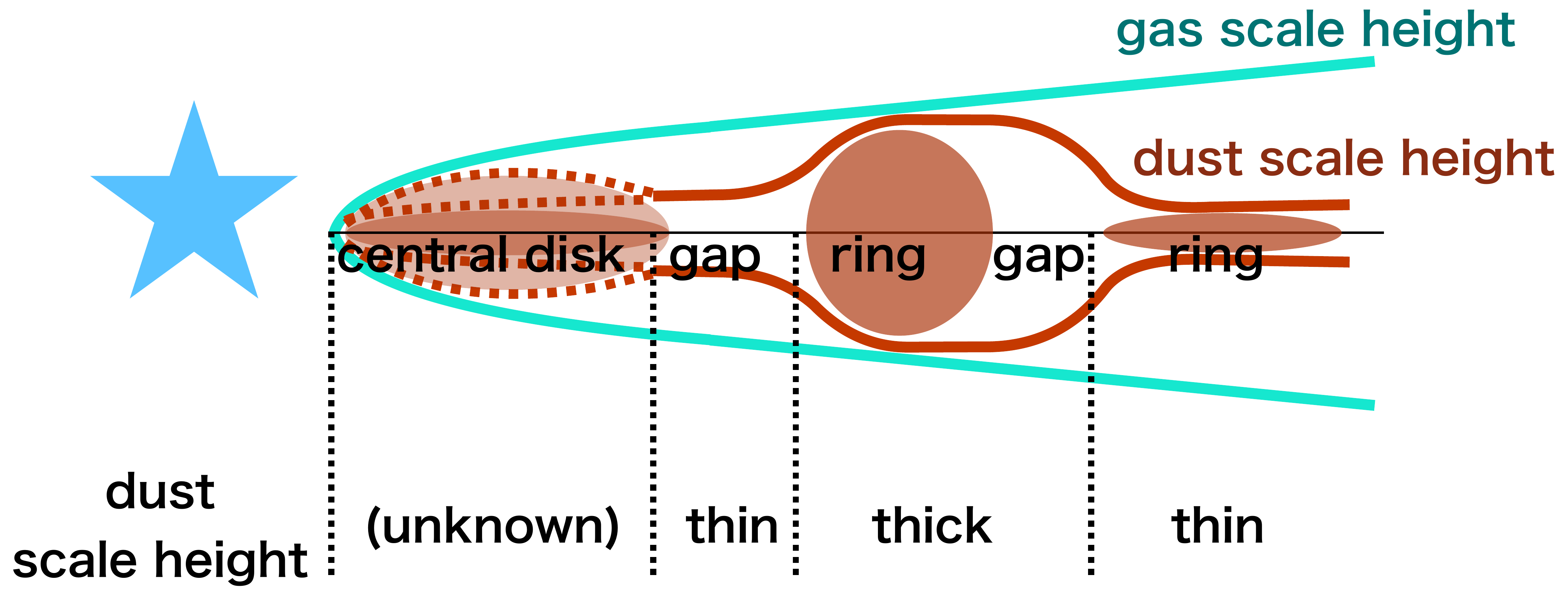}
      \caption{
	Schematic view of the dust scale height for the entire region.
	The scale height in the gaps is taken from \citet{Ohashi2019}, and the scale height in the rings is from our study.
	The dust scale height of the \textcolor{black}{central} disk remains unknown.
}
      \label{scalehight_entire}
  \end{center}
\end{figure}

We discuss the dust scale height for the entire \textcolor{black}{region} of the protoplanetary disk.
Figure \ref{scalehight_entire} shows a schematic illustration of the dust scale height of HD 163296.
\citet{Ohashi2019} determine the dust scale height at the gaps by using the polarized intensity of this object.
They concluded that \textcolor{black}{the dust scale height is less than one-third of the gas scale height at the gap at 48 au, and the dust scale height is two-third of the gas scale height at the gap at 86 au.}
Taken together with the results of this study, we can draw the entire picture of the radial distribution of the dust scale height except for the central disk.
From the inner side to the outer side, the scale height is unknown \textcolor{black}{at} the central disk, thin at the 48 au gap, thick at the 68 au ring and the 87 au gap, and thin at the 100 au ring.

We discuss whether the radial variation of the dust scale height can be explained by a simple dead zone model \citep{Gammie1996, Okuzumi2016, Ueda2019}.
The dead zone exists in the inner part of the protoplanetary disk because the low ionization ratio in the inner part of the disk suppresses the magneto-rotational instability.
The dust is settled in the dead zone due to weak gas turbulence.
Therefore, the dust scale height is smaller in the dead zone and larger outside the dead zone.
As shown in Figure \ref{scalehight_entire}, however, the dust scale height of HD 163296 varies complexly in the radial direction and cannot be explained by a simple disk model with a dead zone.
This suggests that there are other factors that cause the change of the dust scale height, e.g., the radial change of the dust size or the turbulence.

We compare our dust scale height model with an infrared polarization observation.
\citet{Muro-Arena2018} made a 3D disk model of HD 163296 using millimeter dust continuum observation using ALMA and infrared polarized scattered light using SPHERE.
They reported that the polarized infrared light is observed only \textcolor{black}{at} the inner ring at 68 au and is not observed \textcolor{black}{at} the outer ring at 100 au.
This result can be interpreted that the outer ring is in the shadow of the inner ring, which is consistent with our results \textcolor{black}{of the dust scale height.}

\subsection{$\alpha / \mathrm{St}$} \label{subsecalphaSt}

We estimate $\alpha / \mathrm{St}$ from the dust scale height.
By assuming a balance between vertical settling and turbulent diffusion, the dust scale height is written as \citep[e.g.,][]{Dubrulle1995, Youdin2007}
\begin{equation}
 	h_{d} = \left( 1 + \frac{\mathrm{St}}{\alpha} \frac{1 + 2  \mathrm{St}}{1+  \mathrm{St}}\right)^{-1/2} h_{g},
	\label{scale_height}
\end{equation}
where $h$ is scale height, and the subscript $g$ and $d$ indicate gas and dust, respectively.
This equation can be approximated as 
\begin{equation}
 	h_{d} = \left( 1 + \frac{ \mathrm{St}}{\alpha}\right)^{-1/2} h_{g},
	\label{scale_height_simple}
\end{equation}
if $\mathrm{St} \ll 1$.
Using the $f_{\mathrm{set}}$ constrained in section \ref{subsec3.3}, we derive \textcolor{black}{$\alpha/\mathrm{St} > 2.4 $ for the inner ring and $\alpha/\mathrm{St} < 1.1 \times10^{-2}$ for the outer ring.}
\textcolor{black}{We note that equation (\ref{scale_height_simple}) may not be appropriate if $h_d \approx h_g$.}

We discuss the consistency with measurements in other studies.
\citet{Dullemond2018} and \citet{Rosotti2020} estimated $\alpha / \mathrm{St}$ by the radial diffusion of dust due to the gas turbulence.
\citet{Dullemond2018} assumed ring formation by dust traps at gas bumps and derived the following relationship between dust ring width $w_d$ and gas ring width $w_g$.
\begin{equation}
 	w_{d} = \sqrt{ \frac{\alpha}{\mathrm{St}} } w_{g}.
	\label{width_gas_dust}
\end{equation}
\citet{Dullemond2018} estimated the dust ring width from dust continuum emission, and \citet{Rosotti2020} estimated the gas ring width from the deviation of gas rotation velocity.
Their results are $\alpha/ \mathrm{St} = 0.23$ for the inner ring and $\alpha/ \mathrm{St} = 0.04$ for the outer ring.
\textcolor{black}{We note that \citet{Rosotti2020} and our \textcolor{black}{results} show the same tendency that the inner ring is more diffuse than the outer ring.}
\textcolor{black}{However,} their results are inconsistent with our \textcolor{black}{results.}
There are two possible reasons of this disagreement.
One is that the turbulence is not isotropic, i.e., the turbulence is different between radial and vertical directions.
The other is that the ring formation mechanism is different from that assumed by \citet{Rosotti2020}.
\citet{Rosotti2020} assume that ring formation is due to gas bumps.
In other scenarios, however, the dust ring width and gas ring width do not necessarily satisfy the relationship of equation (\ref{width_gas_dust}).
We discuss other ring formation mechanisms in section \ref{subsec:ring}.

\subsection{dust size} \label{dustsize}
In this section, we constrain the dust size based on the $\alpha / \mathrm{St}$.
\textcolor{black}{In section} \ref{subsecalphaSt}, we obtained $\alpha / \mathrm{St}$ based on $f_{\mathrm{set}}$.
The Stokes number is expressed as
\begin{equation} \label{stokes_number}
    \mathrm{St} = \frac{\pi \rho_{\mathrm{mat}} a_{\mathrm{dust}}}{2 \Sigma_{\mathrm{gas}}},
\end{equation}
where $\rho_{\mathrm{mat}}$ is the dust material density, $a_{\mathrm{dust}}$ is the dust radius, and $\Sigma_{\mathrm{gas}}$ is the gas surface density.
Therefore, we can constrain the dust size from $\alpha / \mathrm{St}$ obtained \textcolor{black}{in section} \ref{subsecalphaSt} and $\alpha$, $\Sigma_{\mathrm{gas}}$, and $\rho_{\mathrm{mat}}$.
We assume that the dust is compact and icy ($\rho_{\mathrm{mat}} = 1.0\ \mathrm{g/cm^3}$),
and the gas surface density $\Sigma_{\mathrm{gas}} = 21.1\ \mathrm{g/cm^2}$ for the inner ring and $\Sigma_{\mathrm{gas}} = 10.0\ \mathrm{g/cm^2}$ for the outer ring by \citet{Rab2020} based on molecular \textcolor{black}{line} observations.
\citet{Flaherty2017} put an upper limit that $\alpha < 3 \times 10^{-3}$, so we calculate the dust radius for the typical three turbulence that $\alpha = 3 \times 10^{-3},\ 1 \times 10^{-3},\ 1 \times 10^{-4}$. 
We show the results in Table \ref{tab:dustsize}.
If we assume $\alpha = 1 \times 10^{-3}$, we obtain \textcolor{black}{$a_{\mathrm{dust}} < 5.6 \times 10^{-3}\ \mathrm{cm}$ for the inner ring and $a_{\mathrm{dust}} > 5.7 \times 10^{-1} \ \mathrm{cm}$ for the outer ring}.

\begin{deluxetable*}{ccccccccc}[htbp]
\tablecaption{\textcolor{black}{Estimation of the dust size at typical gas turbulence\label{tab:dustsize}}}
\tablehead{
\colhead{}     & \colhead{$f_{\mathrm{set}}$} & \colhead{$\alpha/\mathrm{St}$} & \colhead{$\Sigma_{\mathrm{gas}}$} & \colhead{$\alpha$} & \colhead{$a_{\mathrm{dust}}$}\\
\colhead{Ring} & \colhead{}               & \colhead{}             & \colhead{($\mathrm{g\ cm^{-2}}$)}    & \colhead{}         & \colhead{(cm)}\\
\colhead{(1)}  & \colhead{(2)}                & \colhead{(3)}          & \colhead{(4)}                     & \colhead{(5)}      & \colhead{(6)}}
\startdata
           &        &        &         &            $3\times10^{-3}$ & $ < 1.7 \times 10^{-2} $ \\
inner ring & $1.1^{+0.1}_{-0.1}$ & $> 2.4$ & 21.0 & $1\times10^{-3}$ & $ < 5.6 \times 10^{-3} $\\
           &        &        &         &            $1\times10^{-4}$ & $ < 5.6 \times 10^{-4} $\\ \hline
           &        &        &         &            $3\times10^{-3}$ & $ > 1.7 $ \\
outer ring & $>9.5 $& $<1.1 \times 10^{-2}$ & 10.0& $1\times10^{-3}$ & $ > 5.7 \times 10^{-1} $ \\
           &        &        &         &            $1\times10^{-4}$ & $ > 5.7 \times 10^{-2} $   
\enddata
\tablecomments{\textcolor{black}{(2) Settling parameters we constrain in section \ref{subsec3.3}. (3) $\alpha / \mathrm{St}$ we constrain in section \ref{subsecalphaSt}. (4) The gas surface density at the peak position of the ring \citep{Rab2020}. (5) The gas turbulence. We assume three typical parameters based on the upper limit from \citet{Flaherty2017}. (6) The dust size calculated from the left parameters.}}
\end{deluxetable*}

Since \citet{Flaherty2017} put an upper limit that $\alpha < 3 \times 10^{-3}$, we can constrain the dust size of the inner ring without any assumption of turbulence.
As shown in Table \ref{tab:dustsize}, the dust size in the inner ring is \textcolor{black}{$a_{\mathrm{dust}} < 1.7 \times 10^{-2}\ \mathrm{cm}$}.
The result rules out \textcolor{black}{models that} the midplane dust \textcolor{black}{is centimeter pebble}.
\textcolor{black}{Note that \citet{Flaherty2017} assumed that $\alpha$ is constant throughout the disk, but this may not be the case.}

If the turbulence is the same in the two rings, we find that the dust size is larger in the outer ring than in the inner \textcolor{black}{ring}.
\citet{Guidi2016} found a trend of the dust size from the spectral index.
They concluded that the dust size is decreasing toward the outside from the increasing trend of the spectral index, while the spatial resolution is not high enough to resolve the rings.
\citet{Dent2019HD163296polarization} obtained specrtal index, $\alpha_{\mathrm{SED}}$, between Band 6 and Band 7 (0.87 mm) from high resolution observations that spatially resolves the ring.
\textcolor{black}{
\citet{Dent2019HD163296polarization} obtained that $\alpha_{\mathrm{SED}}$ in the rings is smaller than that in the gaps, and $\alpha_{\mathrm{SED}}$ in the inner ring ($\alpha_{\mathrm{SED}}=2.2$) is larger than that in the outer ring ($\alpha_{\mathrm{SED}}=1.8$).
}
Since the systematic uncertainty is large, $\alpha_{\mathrm{SED}}$ for total disk flux differs among \citet{Pinilla2014sed} ($2.73 \pm 0.44$), \citet{Guidi2016}, \citet{Notsu2019HD163296} ($2.7$), and \citet{Dent2019HD163296polarization} ($2.1 \pm 0.3$).
The value of $\alpha_{\mathrm{SED}}$ itself may not be accurate, but we can consider that the radial trend of $\alpha_{\mathrm{SED}}$ is correct.
\textcolor{black}{If we assume that this disk is optically thin, the dust is smaller in the inner ring than in the outer ring.}
This result is consistent with our result \textcolor{black}{assuming the same turbulence between the two rings.}
We do not rule out the possibility that the gas turbulence is different between the inner and outer rings.
\textcolor{black}{We note that if the outer ring is optically thick, $\alpha_{\mathrm{SED}}$ becomes smaller in the outer ring, and the azimuthal variation of the intensity is small.}

\citet{Ohashi2019} estimated the \textcolor{black}{maximum} dust size at the dust gap to be 140 $\mathrm{\mu}$m, based on the polarization fraction of ALMA Band 7 (0.87 mm) dust continuum.
This result is smaller than the dust size of the outer ring shown in Table \ref{tab:dustsize}.
The dust size estimated from the polarization observation is just the dust size in the gap, and we cannot directly compare it with the dust size in the ring.
As shown in Table \ref{tab:dustsize}, however, it is consistent to assume that the dust size in the inner ring is the same as that in the gaps. 

\subsection{turbulence} \label{turbulence_subsec}
In this section, we discuss the possible range of $\alpha$ and $\mathrm{St}$.
Since our method cannot solve for the degeneracy of $\alpha$ and $\mathrm{St}$ from $\alpha / \mathrm{St}$, we estimate $\alpha$ under the following two assumptions: one is that the dust size is \textcolor{black}{limited by the collisional} fragmentation due to the turbulence and the other is that the dust size is fixed to be 1 mm.

First, we consider the case where the dust size is \textcolor{black}{limited} by the collisional fragmentation of dust grains induced by gas turbulence.
In this case, $\alpha$ can be written as \citep{Birnstiel2012,Rosotti2020}
\begin{equation}
    \alpha_{\mathrm{frag}} = \sqrt{ \frac{1}{3} \frac{v_{\mathrm{frag}}^2}{c_{\mathrm{s}}^2} \left( \frac{\alpha}{\mathrm{St}} \right) },
\end{equation}
where $v_{\mathrm{frag}}$ is the fragmentation velocity and $c_{\mathrm{s}}$ is the sound speed.
We obtain \textcolor{black}{$\alpha_{\mathrm{frag}} > 3.0 \times 10^{-2}$} for the inner ring and \textcolor{black}{$\alpha_{\mathrm{frag}} < 2.1 \times 10^{-3}$} for the outer ring when we assume the fragmentation velocity to be $v_{\mathrm{frag}} = 10\ \mathrm{m/s}$.
The dust size may not be determined by the fragmentation limit by the turbulence, but if the turbulence is larger than $\alpha_{\mathrm{frag}}$, the dust is destroyed and becomes smaller.
Therefore, $\alpha_{\mathrm{frag}}$ is the upper limit of the turbulence.

Next, we consider the turbulence when the dust size is 1 mm.
As we discuss in \textcolor{black}{section \ref{dustsize}}, the dust size may not be the same for the two rings, but as an example, we calculate the turbulence $\alpha_{1\ \mathrm{mm}}$ when the dust size $a_{\mathrm{dust}} = 1\ \mathrm{mm}$ for the two rings.
The $\mathrm{St}$ is determined as equation (\ref{stokes_number}), and we assume that the gas surface density and dust material density are the same as those in \textcolor{black}{section \ref{dustsize}}.
Under these assumptions, we obtain \textcolor{black}{$\alpha_{\mathrm{1\ mm}} > 1.8 \times 10^{-2}$} for the inner ring and \textcolor{black}{$\alpha_{\mathrm{1\ mm}} < 1.8 \times 10^{-4}$} for the outer ring.

\begin{deluxetable}{cccccc}[htbp]
\tablecaption{\textcolor{black}{turbulence $\alpha$ under certain assumptions}\label{tab:alpha}}
\tablehead{
\colhead{Ring} & \colhead{$\alpha/\mathrm{St}$} & \colhead{$\alpha_{\mathrm{frag}}$}  & \colhead{$\alpha_{\mathrm{1\ mm}}$} \\
\colhead{(1)}  & \colhead{(2)}                & \colhead{(3)}          & \colhead{(4)}                   
}
\startdata
inner ring & $> 2.4$                 & $ > 3.0 \times 10^{-2}$  & $ > 1.8 \times 10^{-2}$\\
outer ring & $ < 1.1 \times 10^{-2}$ & $ < 2.1 \times 10^{-3}$  & $ < 1.8 \times 10^{-4}$\\
\enddata
\tablecomments{\textcolor{black}{(3) Turbulence $\alpha$ where the dust size is limited by the collisional fragmentation due to the turbulence. (4) Turbulence $\alpha$ when the dust size is 1 mm.}}

\end{deluxetable}

\begin{figure}[htbp]
    \begin{center}
      \includegraphics[width=9cm]{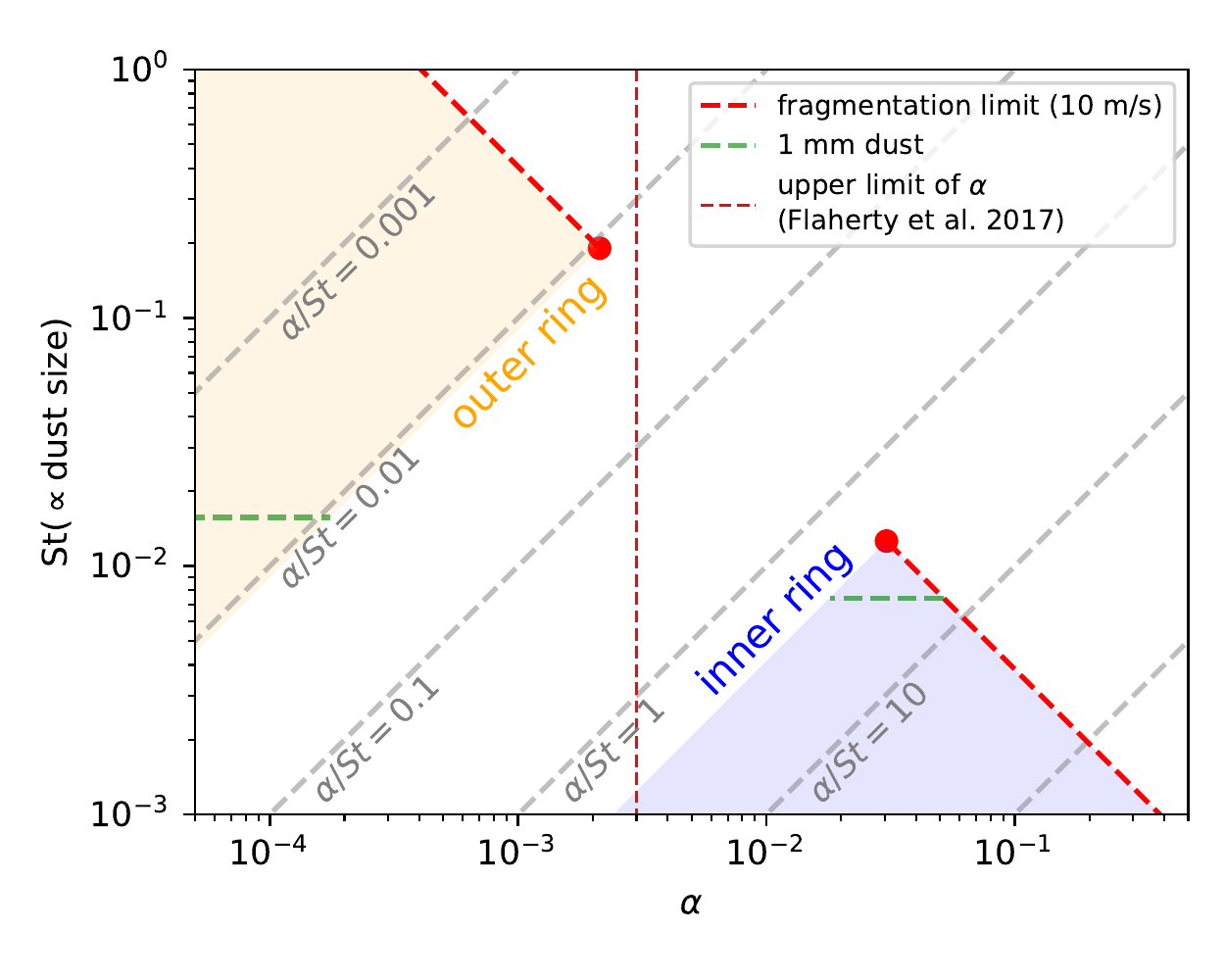}
      \caption{
      \textcolor{black}{The possible range of $\alpha$ and $\mathrm{St}$. The blue and orange areas represent the possible range of $\alpha$ and $\mathrm{St}$ for the inner and outer rings, respectively.}
      The red dashed line represents the fragmentation limit, and the blue and orange lines represent the limit of $\alpha/\mathrm{St}$.
      The brown dashed line represents the upper limit of $\alpha$ from the observation of line broadening \citep{Flaherty2017}.}
      \label{alpha_fig}
  \end{center}
\end{figure}

We summarize the results in Table \ref{tab:alpha} and Figure \ref{alpha_fig}.
\textcolor{black}{
The constraint of $\alpha_{\mathrm{frag}}$ in this study at the inner ring is larger than the observational upper limit of $\alpha$ in \citet{Flaherty2017}.
There are three possible explanations for this inconsistency.
First, there may be a mechanism that keeps the dust size small in the inner ring.
For example, the fragmentation velocity may be smaller than the assumption in this study due to dust sintering \citep{Okuzumi2016}.
Second, the turbulence may vary in the radial direction.
\citet{Flaherty2017} constrained $\alpha$ throughout the disk from gas emission line observations, and this value corresponds to the intensity-weighted average of $\alpha$.
Since the gas disk is larger than the dust ring, low turbulence in the outer side results in a small $\alpha$ throughout the disk.
Third, the $\alpha$ of the dust diffusion we estimate may be different from the $\alpha$ of the gas motion measured by \citet{Flaherty2017}.
\citet{Pinilla2020arXiv} discuss the case where $\alpha$ for gas evolution, turbulent velocity, vertical diffusion, and radial diffusion take different values.}

\textcolor{black}{
\citet{Liu2018} constrained $\alpha$ from hydrodynamic simulations assuming gap opening by the planet.
They showed that if $\alpha$ is small ($\leq 10^{-4}$) in the inner disk and large ($\geq 7.5 \times 10^{-3}$)  in the outer disk, they can reproduce both dust and gas observations.
However, our study shows that $\alpha/\mathrm{St}$ is larger at the inner ring than at the outer ring, which is different from \citet{Liu2018}.
This disagreement indicates that it is difficult to reproduce the gas and dust profiles with the dust trap alone.
The ring formation mechanism will be discussed in section \ref{subsec:ring}.
}

\subsection{ring formation mechanism} \label{subsec:ring}
In this section, we discuss the ring formation process based on the dust scale height, which we estimate in this study.
Previously, several \textcolor{black}{ring formation mechanisms} have been proposed, and we consider the following four mechanisms; 1. accumulation by gas gaps induced by planets, 2. enhanced fragmentation of dust grains by sintering at snowlines, 3. dust accumulation at \textcolor{black}{the dead zone outer edge}, and 4. secular gravitational instability (SGI).
This study shows that the settling conditions of the two rings are different.
Therefore, we consider ring formation mechanisms of each ring independently.

A possible mechanism is dust accumulation by the pressure gaps induced by planets
\citep{Lin1979, Goldreich1980, Duffell2013, Fung2014, Kanagawa2015}.
\citet{Zhu2012} show that large dust accumulates on the outside of the pressure gap, \textcolor{black}{while} small dust couples with gas and flows into the inside of the pressure gap.
This process is called dust filtration.
Since the rings formed by this process consist of large dust, the dust is settled to the midplane.
Compared to the results of our study, the outer ring is consistent with this process because the dust is settled, while the inner ring is inconsistent with this process because the dust is flared.

Another possible mechanism is the dust sintering \citep{Okuzumi2016, Zhang2015HLTau_pebble}.
The sintering theory predicts \textcolor{black}{that} the rings are consist of small dust fragments.
The inner ring may be explained by the dust sintering because the dust is flared, but the outer ring cannot be explained because the dust is settled, which suggests large dust grains.

Another possible mechanism is the dust accumulation at the \textcolor{black}{dead zone outer edge} \citep{Flock2015, Pinilla2016, Mori2016, Ueda2019}.
The dead zone is the inner region of the disk where the MRI is stabilized, and the gas turbulence is weak.
However, since the dust in the inner ring is not settled, the turbulence is not weak, and the dead zone is not the formation mechanism of the ring.
The outer ring is far from the central star when we consider the dead zone.
We do not rule out the possibility that the central disk is due to the dead zone.

The last possible mechanism we consider is secular gravitational instability
\citep[SGI;][]{Ward2000, Youdin2011, Michikoshi2012, Takahashi2014, Tominaga2019}.
The SGI is stabilized by the turbulence, so the turbulence must be weak for \textcolor{black}{the} instability to develop.
Therefore, the dust is settled in the dust ring formed by the SGI.
From the perspective of the dust scale height, the outer ring is consistent, and the inner ring is inconsistent.

To summarize the above, from the perspective of dust scale height, the formation mechanism of the inner ring is consistent with the dust sintering.
In contrast, the formation mechanism of the outer ring is consistent with dust accumulation by gas gaps induced by planets and secular gravitational instability.
We do not rule out other ring-forming mechanisms, such as dust \textcolor{black}{replenishment} from a planet around the substructure.

\subsection{caveats}
We assume the temperature to be isothermal in the vertical direction and simple power law in the radial direction, but this may be different from the actual temperature.

In this study, we focus on only one object, HD 163296.
This method can be applied to other objects to estimate the dust scale height.
By using high spatial resolution observations, it is possible to limit the dust scale height of other objects.

\section{Conclusion}
We developed a method to estimate the dust scale heights of dust rings of protoplanetary disks from dust continuum \textcolor{black}{images}.
We apply the method to the DSHARP dust continuum \textcolor{black}{image} of HD 163296 \citep{Andrews2018}.
Based on the scale height we constrain, we discuss the rings' physical state and \textcolor{black}{the ring} formation mechanisms.
Our conclusions are as follows.
\begin{enumerate}
  \item The intensity of the \textcolor{black}{dust rings} depends on the dust scale height, especially around the minor axes. 
  Therefore, we can estimate the dust scale height by comparing the intensity along the rings' \textcolor{black}{ridges}. 
  To constrain the dust scale height, we need the following conditions;
  1. dust ring is optically thin, 2. the dust ring width is not much wider than the dust scale height, 3. the inclination is not too small, and 4. the dust ring is spatially resolved.
  
  \item We constrain the dust scale height of HD 163296.
  The dust scale height is different between the inner ring at 68 au and the outer ring at 100 au.
  We obtain \textcolor{black}{$f_{\mathrm{set}} = 1.1^{+0.1}_{-0.1}$} for the inner ring and \textcolor{black}{$f_{\mathrm{set}} > 9.5$} for the outer ring \textcolor{black}{with $3 \sigma$ uncertainties}. 
  This infers that the dust is flared in the inner ring, \textcolor{black}{and settled} in the outer ring.
  
  \item We discuss the dust scale height for the entire region of this object by combining this study, which constrains that of the dust ring, and \citet{Ohashi2019}, who constrains that of the dust gap.
  The dust scale height varies complexly in the radial direction, and a simple dead zone model cannot explain the variation of the dust scale height.
  
  \item Based on the dust scale height, we estimate $\alpha / \mathrm{St}$.
  We obtain \textcolor{black}{$\alpha /  \mathrm{St} > 2.4$} in the inner ring, while \textcolor{black}{$\alpha / \mathrm{St} < 1.1 \times 10^{-2}$} in the outer \textcolor{black}{ring.}
  This results show that the turbulence is stronger or the dust is smaller at the inner ring than at the outer ring.

  \item We constrain the dust size based on the derived $\alpha / \mathrm{St}$.
  If we assume the turbulence is the same in the two rings, we find that the dust size is larger in the outer ring than in the inner ring.
  We constrain the dust size in the inner ring \textcolor{black}{$a_{\mathrm{dust}} < 1.7 \times 10^{-2}\ \mathrm{cm}$} by using the upper limit of the gas turbulence by \citet{Flaherty2017}.
  
  \item We discuss the possible range of $\alpha$ and $\mathrm{St}$.
  We assume the fragmentation limit and obtain \textcolor{black}{that $\alpha_{\mathrm{frag}} > 3.0 \times 10^{-2}$} for the inner ring and \textcolor{black}{$\alpha_{\mathrm{frag}} < 2.1 \times 10^{-3}$} for the outer \textcolor{black}{ring.}
  The $\alpha_{\mathrm{frag}}$ in the inner ring is larger than the upper limit of $\alpha$ from the line broadening that $\alpha < 3 \times 10 ^{-3}$ \citep{Flaherty2017}.
  We need other mechanisms to keep the dust small for the consistency of our result and the constraints from the line broadening \citep{Flaherty2017}.
  
  \item We discuss the dust ring formation mechanisms from the perspective of the dust scale height.
  \textcolor{black}{The formation mechanism of the inner ring is consistent with enhanced dust fragmentation by dust sintering, while that of the outer ring is consistent with dust accumulation by planet inducing gas gap or secular gravitational instability.}
\end{enumerate}

\textcolor{black}{The authors thank the anonymous referee for their helpful comments.}
The authors thank T. Tsukagoshi and T. Ueda for fruitful discussion and comments.
This project is first started with the dust continuum data of \citet{Isella2016}, which is kindly provided by A. Isella.
This work was supported by JSPS KAKENHI Grant Numbers 18K13590 and 19H05088. 

\appendix
\section{effect of vertical temperature gradient} \label{appendixA}

In this paper, we consider vertically isothermal disks.
In reality, the surface layer, which is irradiated by the central star, is hotter than the midplane.
We discuss how the vertical temperature structure affects the observed images.
Note that we ignore the effect of accretion heating because we focus on the outer part of the disk \citep{Chiang1997, Balbus_Hawley1998_accretion_disk}.

The dust temperature below one gas scale height is almost the same as that at the midplane \citep{DAlessio1998}.
Most of the dust locates where the height from the midplane is less than one gas scale height even if the dust is flared.
Therefore, the hot upper layer hardly affects the dust continuum image, and it is reasonable to approximate that the disk is vertically isothermal.

The vertical temperature gradient can affect the observational image if the temperature varies significantly near the midplane.
If the disk is optically thick, the contribution of radiation from the hot upper layer is large.
Since the optical depth is larger on the major axis than on the minor axis, we observe radiation from higher layers on the major axis than the minor axis.
Therefore, we expect the dust ring to be brighter on the major axis than on the minor axis.
This trend of the azimuthal variation is the same as the case of the optically thin and geometrically thick ring discussed in section \ref{subsec2.2}.
We tested whether the above two conditions are confusing.

Even though the vertical temperature variation around the midplane is believed to be small, we demonstrate the case where the disk is optically thick and have a strong vertical temperature gradient even around the midplane.
We performed additional radiative transfer simulations with the temperature model as
\begin{eqnarray}
T(z) = \left\{
\begin{array}{ll}
T_a + (T_m - T_a)[\cos{\frac{\pi z}{2 z_q}}]^{2\delta} \ \ & \mathrm{if}\ z < z_q\\
T_a         & \mathrm{if}\ z > z_q,
\end{array}
\right.
\end{eqnarray}
where $T_m = 21.6\ \mathrm{K}$, $T_a = 77.8\ \mathrm{K}$, $\delta = 2.16$ and $z_q = 25.2\ \mathrm{au}$.
These equations are based on the temperature model in \citet{Rosenfeld2013} at 100 au.
We assume that the dust is radially isothermal to focus on the effect of the vertical temperature gradient.
This model leads to the temperature at one gas scale height being almost twice the midplane temperature, which may make some difference in the final images.
The gas scale height is 6.6 au at $r=100\ \mathrm{au}$ based on the midplane temperature.
We assume rings whose ring width, $\sigma_w$, is 10 au and optical depth, $\kappa \Sigma_d$, is 5.
\begin{figure}[htbp]
    \begin{center}
      \includegraphics[width=12cm]{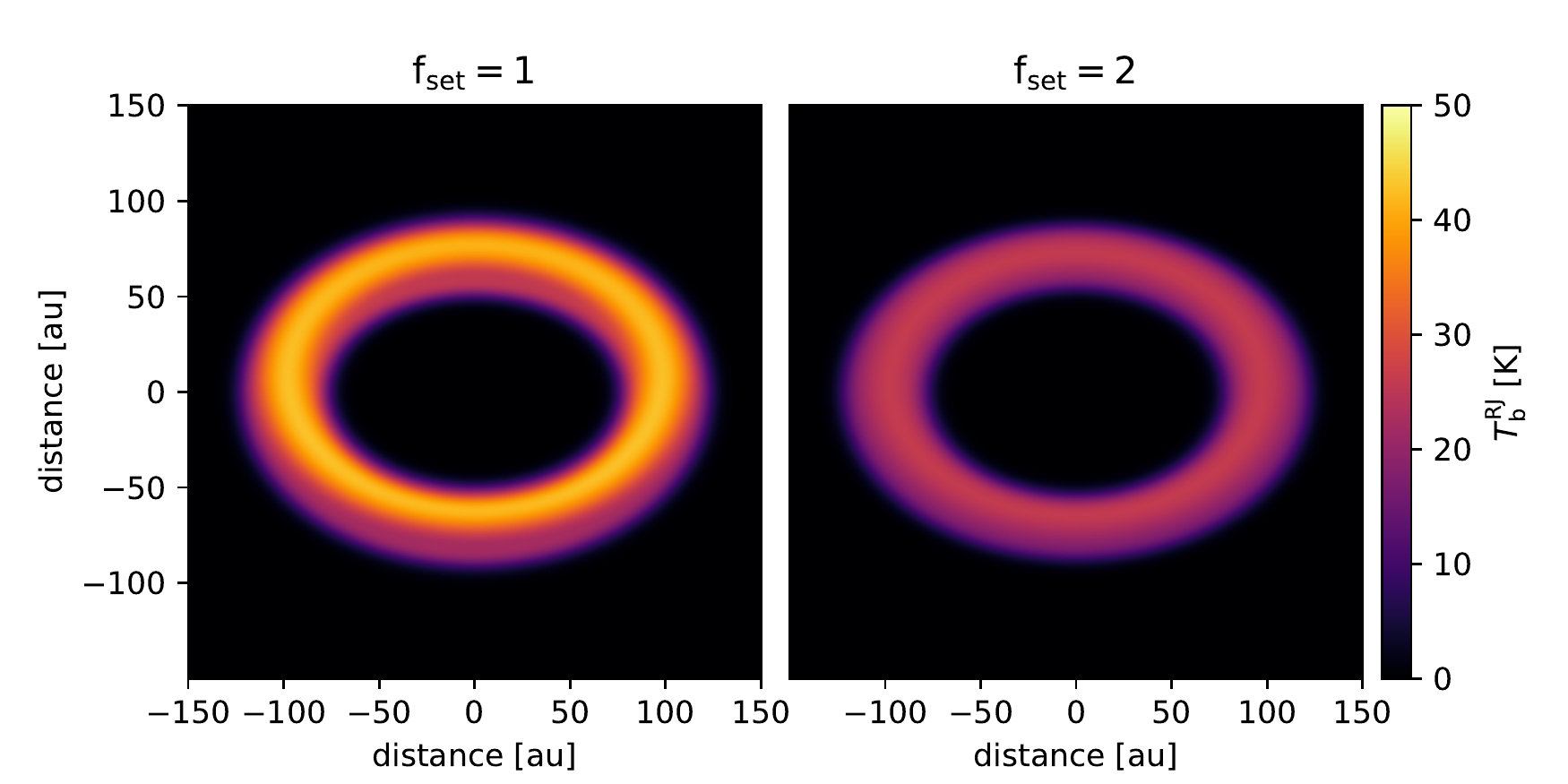}
      \caption{
      Images that demonstrate the effect of the vertical temperature gradient.
    We use the temperature model of \citet{Rosenfeld2013}.
    We assume rings whose ring width, $\sigma_w$, is 10 au and optical depth, $\kappa \Sigma_d$, is 5.
    The left image shows the result for $f_{\mathrm{set}}=1$, and the right image shows the result for $f_{\mathrm{set}}=2$.}
      \label{vertical_temp_image}
  \end{center}
\end{figure}

\begin{figure}[htbp]
    \begin{center}
      \includegraphics[width=8cm]{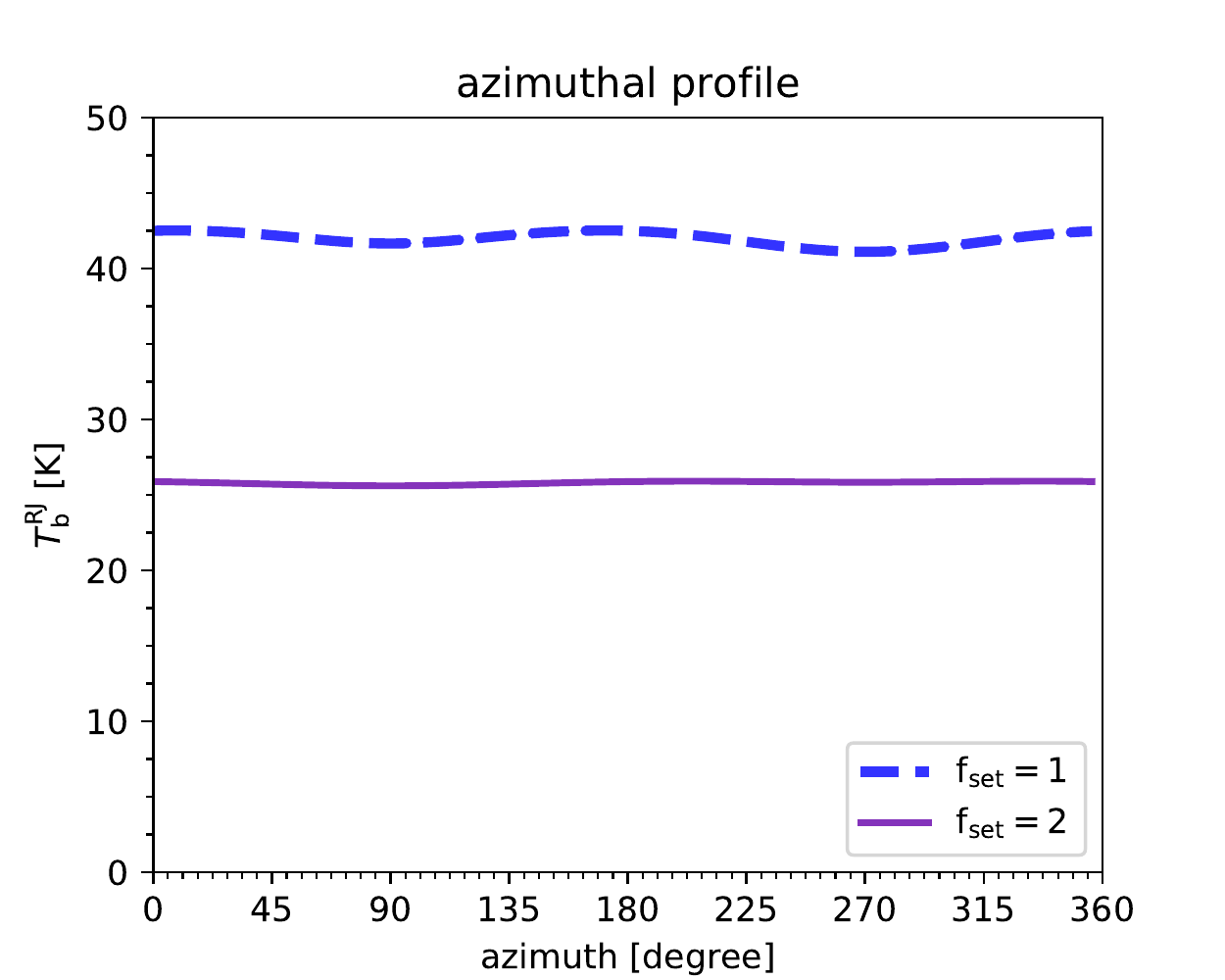}
      \caption{
      Intensity along the ridges in Figure \ref{vertical_temp_image}.
      }
      \label{vertical_temp_degreeBT}
  \end{center}
\end{figure}
Figure \ref{vertical_temp_image} shows the simulation results for $f_{\mathrm{set}}=1$ and 2, and Figure \ref{vertical_temp_degreeBT} shows the intensity along the ridge. 
The intensity is larger than Figure \ref{model_image} if $f_{\mathrm{set}}=1$ because the temperature at the $\tau=1$ surface is higher than that at the midplane.
On the other hand, the intensity is hardly affected by the vertical temperature gradient if $f_{\mathrm{set}}=2$ because the temperature on the $\tau=1$ surface is almost the same as that at the midplane.
The intensity along the ridge is slightly brighter on the major axis than on the minor axis for $f_{\mathrm{set}}=1$ and almost flat for $f_{\mathrm{set}}=2$.
In any case, this pattern can not be confused with the pattern for optically thin and geometrically thick case.

\section{effect of beam smearing} \label{appendix:beam}
We discuss the requirements of the spatial resolution to constrain the dust scale height.
We assume that the observed image is smoothed by an ellipsoidal Gaussian function, as shown in equation (\ref{intensity_general}).
The beam smearing decreases the intensity along the ring's ridge, and can make azimuthal variations of the intensity, especially if the beam is elongated.

\begin{figure*}[htbp]
    \begin{center}
      \includegraphics[width=20cm]{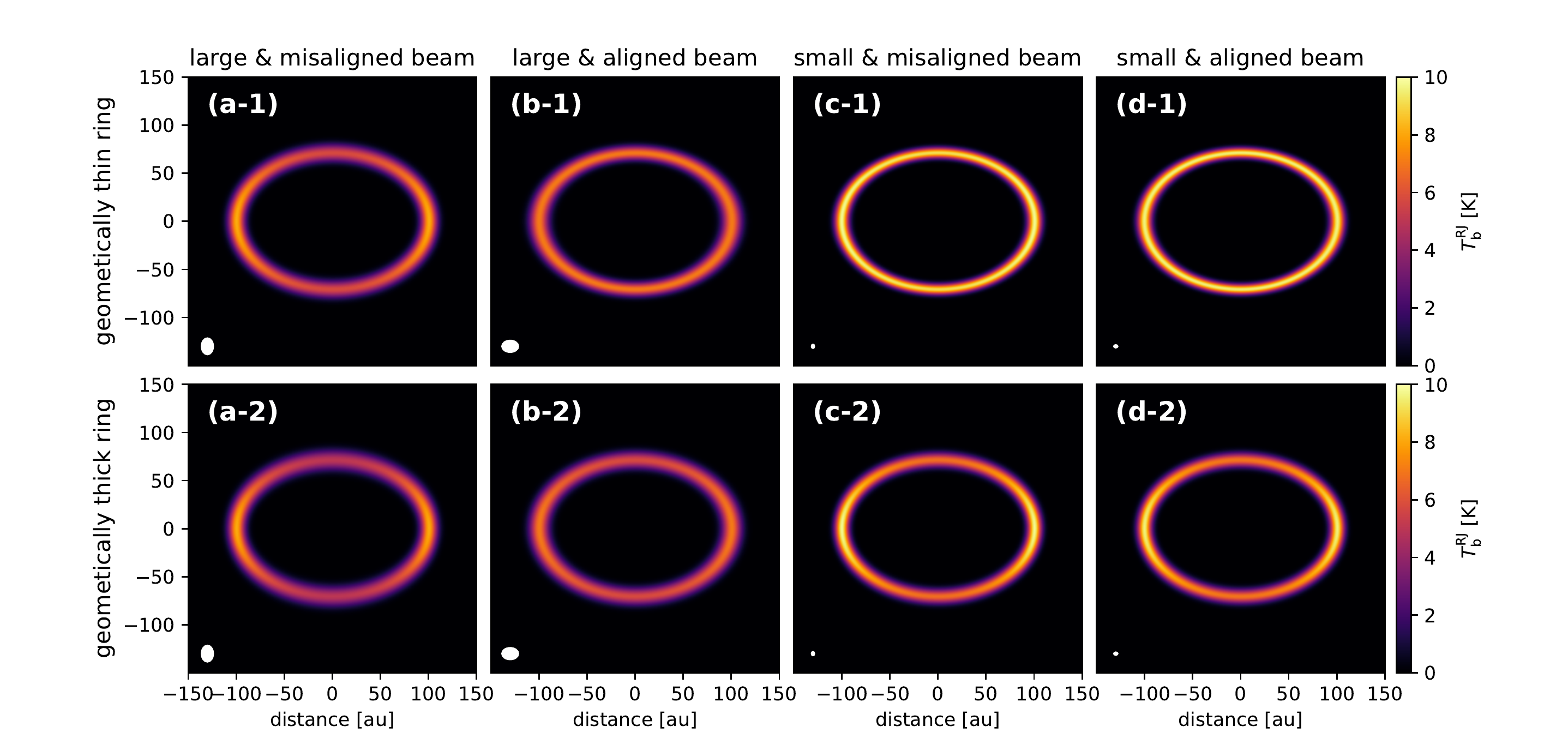}
      \caption{
      Images after beam smearing.
(a-1), (b-1), (c-1), and (d-1) (i.e. the upper 4 panels) are the geometrically thin ring models, and (a-2), (b-2), (c-2), and (d-2) (i.e. the lower 4 panels) are the geometrically thick ring models.
(a-1), (a-2), (b-1), and (b-2) (i.e. the left 4 panels) show images convolved with large beams ($\sigma_{\mathrm{beam}} = 5 \sqrt{2}\ \mathrm{au}$ along the major axis and $\sigma_{\mathrm{beam}} = 5\ \mathrm{au}$ along the minor axis), and (a-2), (b-2), (c-2), and (d-2) (i.e. the right 4 panels) show images convolved with small beams ($\sigma_{\mathrm{beam}} = 1.5\ \mathrm{au}$ along the major axis and $\sigma_{\mathrm{beam}} = 1.5/\sqrt{2}\ \mathrm{au}$ along the minor axis).
The beams' and the rings' position angles are aligned in (b-1), (b-2), (d-1) and (d-2), and those are perpendicular in (a-1), (a-2), (c-1) and (c-1).
}
      \label{beam_image}
  \end{center}
\end{figure*}

\begin{figure}[htbp]
    \begin{center}
      \includegraphics[width=15cm]{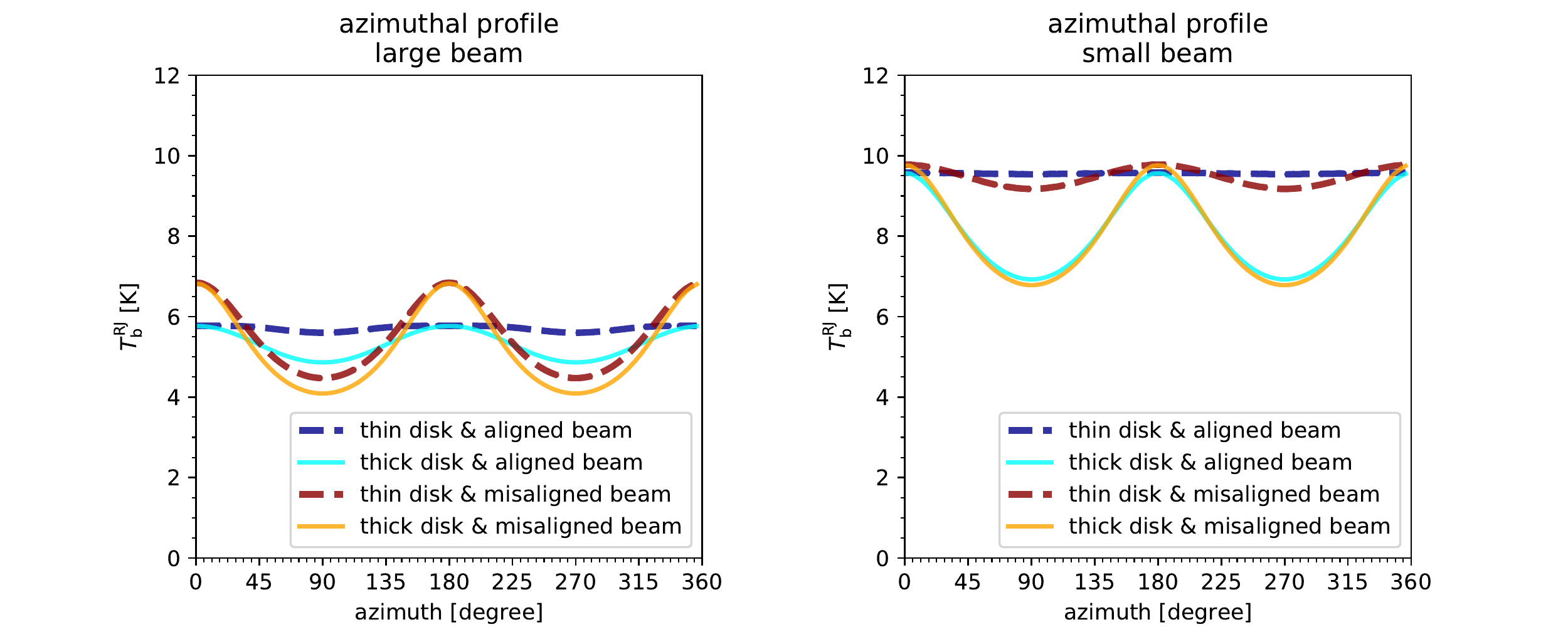}
      \caption{
      Intensity along the ridges in Figure \ref{beam_image}.
      }
      \label{beam_degreeBT}
  \end{center}
\end{figure}

We convolve the model images with beams and discuss the requirements to distinguish the difference of the dust scale height.
Figure \ref{beam_image} shows convolved images, and Figure \ref{beam_degreeBT} shows the intensity of these images along the ridges.
The ring models are the same as those in the middle and right panels of Figure \ref{model_image}.
Figure \ref{beam_image} (a-1), (b-1), (c-1), and (d-1) (i.e. the upper 4 panels) show the geometrically thin ring models, and Figure \ref{beam_image} (a-2), (b-2), (c-2), and (d-2) (i.e. the lower 4 panels) show the geometrically thick ring models.
Figure \ref{beam_image} (a-1), (a-2), (b-1), and (b-2) (i.e. the left 4 panels) show images convolved with large beams ($\sigma_{\mathrm{beam}} = 5 \sqrt{2}\ \mathrm{au}$ along the major axis and $\sigma_{\mathrm{beam}} = 5\ \mathrm{au}$ along the minor axis), and Figure \ref{beam_image} (c-1), (c-2), (d-1), and (d-2) (i.e. the right 4 panels) show images convolved with small beams ($\sigma_{\mathrm{beam}} = 1.5\ \mathrm{au}$ along the major axis and $\sigma_{\mathrm{beam}} = 1.5/\sqrt{2}\ \mathrm{au}$ along the minor axis).
The beams' and the rings' major axes are aligned in Figure \ref{beam_image} (b-1), (b-2), (d-1), and (d-2), and the deprojected beams are perfect circles.
The beams' and the rings' major axes are perpendicular in Figure \ref{beam_image} (a-1), (a-2), (c-1), and (c-2), and the major axes of the deprojected beams are twice as large as the minor axes of them.

Figure \ref{beam_image} (a-1) shows that the beam smearing makes the azimuthal intensity variations even though the ring is geometrically thin.
Therefore, we cannot simply conclude that the dust scale height is large from the presence of the azimuthal intensity variation if the beam size is larger than the scale height.
The azimuthal intensity variation in the left panel of Figure \ref{beam_degreeBT} is almost the same whether the dust scale height is large or small, and we cannot constrain the dust scale height under such a beam.

Figure \ref{beam_image} (b-1) shows that the beam does not make the azimuthal intensity variation even though the beam size is the same as that of Figure \ref{beam_image} (a-1).
We can also distinguish differences in the dust scale heights from the azimuthal intensity variation in the right panel of Figure \ref{beam_degreeBT}.

Figure \ref{beam_image} (c-1), (c-2), (d-1), (d-2) show that the beam smearing hardly affect observational images if the beam size is small.
We can clearly distinguish differences in the dust scale heights from the azimuthal intensity variation in the right panel of Figure \ref{beam_degreeBT}.

To summarize the discussion above, a small deprojected beam is desirable, i.e., a small observed beam and aligned position angles of the beam and ring.
Especially if the beam is large and the beam's and ring's position angles are misaligned, the beam smearing can make azimuthal intensity variations even if the ring is geometrically thin.
Since the beam smearing reduces the intensity along the ridge, we should convolve model images with the beam before comparing them with the observation.

\bibliography{doi_citation}
\bibliographystyle{aasjournal}

\end{document}